\documentclass[twocolumn]{aastex62}

\usepackage{amsmath}

\newcommand\Cholla{\emph{Cholla}~}
\newcommand\Chollan{\emph{Cholla}}
\newcommand\Enzo{\emph{Enzo}~}
\newcommand\Enzon{\emph{Enzo}}
\newcommand\Nyx{\emph{Nyx}~}

\newcommand\Ramses{\emph{Ramses}~}

\newcommand\Lya{Lyman-$\alpha$~}
\newcommand\Sim{CHIPS~}
\newcommand\Simn{CHIPS}
\newcommand\skm{\mathrm{s}~\mathrm{km}^{-1}}
\newcommand\HI{\ion{H}{1}~}

\newcommand\HeII{\ion{He}{2}~}

\graphicspath{{./}{figures/}}

\shorttitle{{\it CHOLLA} Simulations of the \Lya Forest}
\shortauthors{Villasenor et al.}

\begin{document}

\title{ Effects of Photoionization and Photoheating on \Lya Forest Properties from \Cholla Cosmological Simulations }

\correspondingauthor{Bruno Villasenor}
\email{brvillas@ucsc.edu}

\author[0000-0002-7460-8129]{Bruno Villasenor}
\affiliation{Department of Astronomy and Astrophysics, University of California,
             Santa Cruz, 1156 High Street, Santa Cruz, CA 95064 USA}

\author[0000-0002-4271-0364]{Brant Robertson}
\affiliation{Department of Astronomy and Astrophysics, University of California,
             Santa Cruz, 1156 High Street, Santa Cruz, CA 95064 USA}

\author[0000-0002-6336-3293]{Piero Madau}
\affiliation{Department of Astronomy and Astrophysics, University of California,
             Santa Cruz, 1156 High Street, Santa Cruz, CA 95064 USA}

\author[0000-0001-9735-7484]{Evan Schneider}
\affiliation{Department of Physics and Astronomy \& Pittsburgh Particle Physics, Astrophysics, and Cosmology Center (PITT PACC), University of Pittsburgh, Pittsburgh, PA 15260, USA}

\begin{abstract}
The density and temperature properties of the intergalactic medium (IGM)
reflect the heating and ionization history during cosmological structure
formation, and are primarily probed by the \Lya forest of neutral hydrogen
absorption features in the observed spectra of background sources
\citep{gunn1965a}. We
present the methodology and initial
results from the \Cholla IGM Photoheating Simulation (\Simn) suite performed with the 
Graphics Process Unit-accelerated \Cholla code to study the
IGM at high, uniform spatial resolution maintained over large volumes.
In this first paper, we examine the IGM structure in \Sim cosmological
simulations that include
IGM uniform photoheating and photoionization models 
where hydrogen reionization completes early \citep{haardt2012a} or by 
redshift $z\sim6$ \citep{puchwein2019a}. Comparing with observations
of the large- and small-scale
\Lya transmitted flux power spectra $P(k)$ at redshifts $2\lesssim z \lesssim5.5$,
the relative agreement of the models depends on scale, with 
the self-consistent \citet{puchwein2019a} IGM photoheating and photoionization
model in good agreement with the flux $P(k)$ at $k\gtrsim0.01~\skm$
at redshifts $2\lesssim z \lesssim 3.5$.
On larger scales the $P(k)$ measurements
increase in amplitude from $z\sim4.6$ to $z\sim2.2$ faster than the models, and
lie in between the model predictions at $2.2\lesssim z \lesssim 4.6$ for 
$k\approx 0.002-0.01~\skm$. We argue the models could improve by changing 
the \HeII photoheating rate associated with active galactic nuclei to
reduce the IGM temperature at $z\sim3$.
At higher redshifts $z\gtrsim4.5$ the observed flux $P(k)$ amplitude 
increases at a rate intermediate between the models, and we argue that for models where hydrogen
reionization completes late ($z\sim5.5-6$) resolving
this disagreement will require inhomogeneous or ``patchy'' reionization.
We then use an additional set of simulations to demonstrate our
results have numerically converged and are not strongly affected by
varying cosmological parameters.
\end{abstract}


\section{Introduction} \label{sec:intro}

The absorption signatures of neutral hydrogen gas provide important
observational probes of cosmological structure formation \citep{gunn1965a}. The
intergalactic medium traces the filamentary structure of the cosmic web,
and the properties of \HI absorption features (the ``\Lya forest'')
reflect the temperature
and density distribution of the medium that originate through
the structure formation process and the photoheating from
ionizing sources \citep[e.g.,][]{madau1999a}. 
This paper presents the first
results from the new \Cholla IGM Photoheating Simulation (\Simn) suite of cosmological
simulations of the \Lya forest performed with the \Cholla code
\citep{schneider2015a}, comparing the statistics of the
simulated \Lya forest calculated using different photoionization
and photoheating histories with the available observational data
at $z\sim2-5$.

The \Lya forest originates in IGM gas that traces the matter density, 
and its properties inform us about 
the relative abundance of baryons
and dark matter, the properties of dark matter including
the matter power spectrum, the metagalactic
radiation field, and the expansion history of the universe
including the role of dark energy
\citep[for a review, see][]{mcquinn2016a}.
The promise of the Lyman-$\alpha$ forest for
constraining the nature of dark matter and dark energy
has in part
motivated the construction of the Dark Energy Spectroscopic
Instrument, which will measure absorption line spectra backlit by quasars at
$z>2.1$ and detect baryon acoustic oscillations 
in the 
cosmic web \citep{desi2016a,desi2016b}. 

Given its critical role as a probe of cosmic structure
formation, the \Lya forest was an early subject of
hydrodynamical cosmological simulations \citep[e.g.,][]{cen1994a,hernquist1996a}.
The prospect of measuring quasar absorption spectra densely sampled on the sky
over large statistical volumes
has led to a resurgence of cosmic web studies in 
the literature \citep[e.g.,][]{lukic2015a,sorini2018a,krolewski2018a}.
Owing to the power of DESI and other
new spectroscopic facilities, the driving focus
of theoretical efforts is to study the
physics that affect the fine details
of the 
forest \citep[e.g.,][]{rorai2017a}.  
These physics
include non-linear effects
\citep{arinyo-i-prats2015a}, environment
\citep{tonnesen2017a}, 
and how the forest evolves to low
redshifts \citep{khaire2019a},
but a consensus is
building that the impact of IGM heating history on
the temperature structure of the
Lyman-$\alpha$ forest is the most
critical to understand in detail \citep[e.g.,][]{hiss2018a}.
The temperature structure affects
most strongly the shape of the
absorption profiles that provide
information about the matter distribution,
and without understanding the thermal
structure of filaments the 
full power of Lyman-$\alpha$ 
absorption line studies cannot
be realized. 

The statistical properties of the Lyman-$\alpha$
forest are primarily measured via the ``transmitted
flux power spectrum'', which probes fluctuations
in the opacity (and therefore density, temperature and velocity field) 
of neutral hydrogen via transmission of flux from
background quasars or galaxies. Similarly to the matter
power spectrum, whose measurements extend to large ($>100$ Mpc comoving)
scales via galaxy spatial correlations, the baryonic
acoustic oscillations are probed by the Lyman-$\alpha$ forest correlations at these large scales as well \citep[e.g.,][]{BOSS_2019}. Structure in the \Lya forest extends down
to scales of $\sim50$ kpc comoving, where thermal pressure
smoothing becomes important \citep{kulkarni2015a}.
Simultaneously
capturing representative volumes while resolving the
relevant spatial scales everywhere in the forest
presents a challenging goal for cosmological simulations.

While dark matter and cosmological structure formation
erect the scaffolding of the cosmic web, 
the temperature structure of the IGM depends on the
competition between heating, radiative cooling, and
adiabatic cooling via universal expansion.
Heating of the IGM predominately occurs via photoheating
from excess energy deposited during the photoionization of
(most importantly) the \HI and \HeII species. Observationally, the
\Lya forest probes redshifts $z<6$ when \HI has mostly been
ionized. Between $4\lesssim z \lesssim 6$, the IGM temperature
declines from a local maximum at the end of \HI reionization
from which the IGM thermal structure inherits residual signatures
\citep{onorbe2017b,gnedin2017a,davies2018a,daloisio2019a,faucher-giguere2020a}. At redshifts $z\lesssim4$, photoheating
from the gradual ionization of \HeII from quasars leads to a maximum
IGM temperature sometime around $z\sim3$ \citep{laplante2017a,schmidt2018a}.
The low-redshift ($2\lesssim z \lesssim4$)
IGM 
is therefore heavily influenced by \HeII reionization \citep{worseck2016a},
and the helium Lyman-{\ensuremath{\alpha}}
forest \citep{laplante2018a}
provides critical information on 
the ionizing flux from quasars
\citep{laplante2017a,schmidt2018a}.

At higher redshifts the ionizing flux
from galaxies becomes increasingly important.
The various
transitions of the hydrogen Lyman series provide
details on the ionization state of the
gas, and can constrain the
post-reionization ionizing background
\citep{davies2018a}.
The
hydrogen reionization process heats
the IGM sufficiently to leave
residual signatures in the structure
of the filaments \citep{daloisio2019a}.
The thermal evolution of the IGM, reflecting
early $z\sim6$ heating from galaxies during H reionization
and late $z\sim2-4$ heating from QSOs during helium reionization,
can therefore be probed through
the Lyman-$\alpha$ forest 
power spectrum \citep{walther2019a}.

By changing the thermal history of the
IGM, the process of cosmic reionization
at $z>6$ couples to the
observed properties of the
Lyman-$\alpha$ forest on small scales. 
Probes of reionization have become
increasingly powerful, including
quasar proximity zones
\citep{eilers2017a} and the
IGM damping wing
\citep{davies2018b},
the high-redshift 
forest and post-reionization IGM
\citep{onorbe2017b,gnedin2017a},
and
Lyman-$\alpha$ transmission spikes
\citep{garaldi2019a,gaikwad2020a}.
The physics of reionization
has driven a host of
cosmological simulation efforts
\citep[e.g.,][]{gnedin2014a,kaurov2014a,kaurov2015a,trac2015a,gnedin2016a,onorbe2017a,doussot2019a},
but more work is required to connect these
simulations to the physics
of the 
IGM at lower redshifts.
Capturing fluctuations in the
metagalactic background
\citep{daloisio2018a} and the
potential impact of rare AGN
\citep{daloisio2017a} require
large volumes ($L\sim100h^{-1}$Mpc),
but simultaneously 
maintaining high spatial resolution in
the IGM is computationally demanding.

\Cholla models the baryionic component on a uniform Cartesian grid, and while other approaches to solve the hydrodynamics can be employed such as SPH or AMR, for which the computational power is commonly focused in solving the high density regions. These approaches present some disadvantages compared to a uniform grid when applied to the study of the IGM. For instance, the gas responsible for the Lyman-$\alpha$ forest is low density gas that spans over most of the volume of the box, making the use of AMR unnecessary and inefficient. Compared to an SPH approach, grid methods exhibit other advantages such as a clearly defined spatial resolution instead of a fixed mass resolution and generally a more accurate treatment of shocks and hydrodynamics. More importantly, a uniform grid achieves a better sampling of the IGM when compared to an AMR or SPH implementation that uses the same computational resources. Although, the advantages of a uniform grid come at a high computational cost if a high resolution is maintained over large volumes.

To this end, our new \Sim simulation suite
uses the Graphics Processing Unit (GPU)-native \Cholla code \citep{schneider2015a,schneider2017a}
to perform high-resolution simulations of the cosmic web
to achieve simultaneously the resolution
required to model the thermal structure of the
Lyman-$\alpha$ forest 
($\Delta x\approx35$ kpc) 
over large volumes ($L\approx75$Mpc). 
We will study how different \HI + \HeII photoionization
and photoheating histories
shape the thermal structure of the IGM.

This first paper presents \Lya forest results for the 
widely-used \citet{haardt2012a} photoionization and
photoheating model, as well as the more recent \citet{puchwein2019a}
implementation 
that has a similar emissivity but for which \HI
reionization completes later ($z\sim6$). Section \ref{sec:methods}
presents our numerical methodology for performing the cosmological simulations
including our new
extensions to the \Cholla code
that enable, self-gravity, dark matter particle integration, and
coupling to the GRACKLE heating and cooling library \citep{smith2017a}.
Section \ref{sec:algorithm} provides a high-level summary of the algorithm used
in our cosmological simulations. Section \ref{sec:validation} presents
several validation tests, including a new validation test for the dual-energy
formalism when modeling cosmological structure formation.
We present the first simulations of the \Sim suite in \S
\ref{sec:simulation_suite}, including the cosmological parameters,
resolution, and box sizes. Our scientific results for the properties
of the IGM are reported in \S \ref{sec:igm_evolution}. We discuss our results in \S
\ref{sec:discussion}, and summarize and conclude in
\S \ref{sec:summary}. Finally, we
 demonstrate the numerical convergence of our results in Appendix
\ref{sec:resolution} and perform a
cosmological parameter study in Appendix \ref{sec:parameter_study}

\section{Methodology} \label{sec:methods}

To simulate the \Lya-forest, we engineered
substantial extensions to the \Cholla code. These
additions included implementing a cosmological 
framework to account for the expansion history 
of the universe (\S \ref{sec:framework}),
including changes to the model of gas dynamics
(\S \ref{sec:gas_dynamics}) and the coordinate
system (\S \ref{sec:coordinates}), and
are discussed below. We briefly review the
\Cholla hydrodynamical integrator (\S \ref{sec:hydro})
and the dual energy formalism (\S \ref{sec:dual_energy})
that allows for accurate evolution of the gas
internal energy in Eulerian cosmological simulations
\citep[e.g.,][]{bryan1995a}.
We present our new implementations of solvers for the
gravitational force
and particle motions in \S \ref{sec:gravity} and
\S \ref{sec:dark_matter_particles}, respectively.
Cooling and heating from
a UV background are
now treated using the GRACKLE library \citep{smith2017a},
and are detailed in \S \ref{sec:cooling} and \S
\ref{sec:uvb_heating}. Adjustments to the time step
calculation to account for particle motions are described
in \S \ref{sec:time_step}. We conclude the review
of our methods
with a summary of the overall algorithm in \S
\ref{sec:algorithm}.

\subsection{Cosmological Framework}
\label{sec:framework}

For cosmological simulations, the gas follows the equations of hydrodynamics in a frame comoving with the expanding universe. To convert from the comoving to the physical system, the scale factor $a$ is introduced and provides a distance transformation between the two systems, with coordinates in the proper system  $\mathbf{r}$
related to comoving coordinates $\mathbf{x}$ by $\mathbf{r} = a \mathbf{x}$. The rate of change of the scale factor corresponds to the expansion rate of the universe  and follows the Friedmann equation given by
\begin{equation}
\label{eqn:friedmann}
    H \, = \, \frac{\dot{a}}{a} \, = \, H_0 \sqrt{ \frac{\Omega_M}{a^3} +  \Omega_{\Lambda} +  \frac{ \Omega_k}{a^2} },
\end{equation}
\noindent
where $H$ is the Hubble parameter that quantifies the expansion rate of the universe $\dot{a}= da/dt$,
and $H_0$, $\Omega_M$, $\Omega_{\Lambda}$ and $\Omega_k$ are the cosmological parameters that correspond to the current expansion rate of the universe and its  matter, dark energy, and curvature content,
respectively. 
Given an initial value of the scale factor $a$, the Friedmann equation provides a relation between the scale factor and cosmic time, and therefore the scale factor can be used as a time-like variable.

\subsubsection{Gas Dynamics }
\label{sec:gas_dynamics}

Consider the hydrodynamical quantities of comoving baryon
density $\rho_b$, proper peculiar velocity $\mathbf{v}$, 
and total specific energy  $E$ in the proper frame.
The relation between comoving and proper densities is $\rho = a^3 \rho_{proper}$. 
In this system, the basic equations of hydrodynamics include the continuity equation
\begin{equation}
\label{eq:continuity}
\frac{\partial \rho_b}{\partial t}  = -\frac{1}{a} \nabla \cdot ( \rho_b \mathbf{v}),
\end{equation}
\noindent
the force-momentum equation
\begin{equation}
\label{eq:momentum}
\frac{\partial a \rho_b \mathbf{v}}{\partial t} = - \nabla \cdot ( \rho_b \mathbf{v} \mathbf{v} ) - \nabla p + \rho_b \mathbf{g},
\end{equation}
\noindent
where the pressure $p$ transforms to the proper pressure by the relation $p=a^3 p_{proper}$ and $\mathbf{g}$ is the gravitational acceleration,
and the energy equation
\begin{align}
\begin{split}
\frac{\partial a^2 \rho_b E}{\partial t} & =  -a \nabla \cdot ( \rho_b \mathbf{v} E + p\mathbf{v} ) + a\rho_b \mathbf{v} \cdot \mathbf{g} \\
& \,\,\,\,\,\, + a \dot{a} [(2-3(\gamma-1))\rho_b e] + a ( \Gamma  - \Lambda )
\label{eq:energy_1},
\end{split}    
\end{align}
\noindent

where $\Gamma$ and  $\Lambda$ correspond to the heating and cooling rates, respectively.

From the specific total energy one can obtain the specific internal energy $e$ in the proper system by subtracting the kinetic energy per unit mass $e = E - v^2/2$.
In Eulerian cosmological simulations where gas often flows supersonically, the above equations can be supplemented
by a dual energy formalism \citep{bryan1995a} in which the internal energy is additionally followed.
The supplemental internal energy is
then used in cells where the computation of the internal energy from the total energy 
is expected to be inaccurate (see \S \ref{sec:dual_energy} for details).  

The evolution of the specific internal energy $e$ is given by 
\begin{equation}
\begin{split}
\frac{\partial a^2 \rho_b e}{\partial t} & = -a \nabla \cdot ( \rho_b \mathbf{v} e ) - a p \nabla \cdot \mathbf{v} \\  
& \,\,\,\,\,\, + a \dot{a} [(2-3(\gamma-1))\rho_b e] + a ( \Gamma  - \Lambda )
\label{eq:internal_energy_1}
\end{split}
\end{equation}
\noindent
The relation between the pressure and the internal energy is given by the equation of state $p = ( \gamma - 1 ) \rho_b e$.  

For simplicity, first we limit the description of the hydrodynamics solver to the adiabatic case ($\Gamma=\Lambda=0$),
and delay a description of the radiative cooling implementation to \S \ref{sec:cooling}.
In the particular case of a $\gamma = 5/3$ gas, the adiabatic energy equations simplify to
\begin{equation}
\begin{aligned}
\frac{\partial a^2 \rho_b E}{\partial t} & =  -a \nabla \cdot ( \rho_b \mathbf{v} E + p\mathbf{v} ) + a\rho_b \mathbf{v} \cdot \mathbf{g},   
\\
\frac{\partial a^2 \rho_b e}{\partial t} & = -a \nabla \cdot ( \rho_b \mathbf{v} e ) - a p \nabla \cdot \mathbf{v}
\label{eq:energy_2}.
\end{aligned}
\end{equation}
\noindent
From Equations \ref{eq:continuity}, \ref{eq:momentum}, and \ref{eq:energy_2} it follows that for a uniform expanding universe, the comoving density $\rho_b$ will remain constant, the peculiar velocity $\mathbf{v}$ will decrease as $a^{-1}$ and the specific energies $E$ and $e$ will decrease as $a^{-2}$.

\subsubsection{Super-comoving Coordinates }
\label{sec:coordinates}

A convenient approach for the implementation of the comoving coordinate system is to define a new set of coordinates that simplify Equations \ref{eq:continuity}, \ref{eq:momentum}, and \ref{eq:energy_2} such that the scale factor $a$ does not explicitly appear.
A detailed description of these ``super-comoving coordinates'' can be found in \cite{Martel+1998} and are used for cosmological simulations in the \Ramses code by \cite{teyssier2002a}. The transformation to the new system of coordinates is given by
\begin{equation}
\begin{aligned}
d\tilde{t}  \equiv H_0 \frac{dt}{a^2}, \;\;\;&\;\;\; \mathbf{\tilde{v}}  \equiv a \frac{\mathbf{v}}{H_0}, \\ 
\tilde{E}  \equiv a^2  \frac{E}{H_0^2}, \;\;\;&\;\;\; \tilde{e}  \equiv a^2  \frac{e}{H_0^2},    \\
\tilde{\rho_b} \equiv \rho_b, \;\;\;&\;\;\;  \tilde{p}  \equiv a^2 \frac{p}{H_0^2} =  ( \gamma - 1 ) \tilde{\rho_b} \tilde{ e}, \\
\tilde{\phi}  \equiv a^2 \frac{\phi}{H_0^2}, \;\;\;&\;\;\; \tilde{\mathbf{g}} = -\nabla \tilde{\phi}  = a^2 \frac{\mathbf{g}}{H_0^2}.
\label{eq:coord_transformation}
\end{aligned}
\end{equation}
\noindent
Throughout we will denote super-comoving variables with a tilde, e.g., $\tilde{\phi}$.
After the transformation to the super-comoving system of coordinates, the equations of adiabatic hydrodynamics for a $\gamma=5/3$ gas can be written as
\begin{align}
\label{eq:hydro_start}
\frac{\partial \tilde{\rho_b}}{\partial \tilde{t}} & = - \nabla \cdot ( \tilde{\rho_b} \mathbf{\tilde{v}}) \\
\label{eq:hydro_momentum}
\frac{\partial \tilde{\rho_b} \mathbf{\tilde{v}}}{\partial \tilde{t}} & = - \nabla \cdot ( \tilde{\rho_b} \mathbf{\tilde{v}} \mathbf{\tilde{v}} ) - \nabla \tilde{p} + \tilde{\rho_b} \mathbf{\tilde{g}} \\
\label{eq:hydro_energy}
\frac{\partial \tilde{\rho_b} \tilde{E}}{\partial \tilde{t}} & = - \nabla \cdot ( \tilde{\rho_b} \mathbf{\tilde{v}} \tilde{E} + \tilde{p} \mathbf{\tilde{v}} ) + \tilde{\rho_b} \mathbf{\tilde{v}} \cdot \mathbf{\tilde{g}}   \\
\frac{\partial \tilde{\rho_b} \tilde{e}}{\partial \tilde{t}} & = - \nabla \cdot ( \tilde{\rho_b} \mathbf{\tilde{v}} \tilde{e} ) -  \tilde{p} \nabla \cdot \mathbf{\tilde{v}}
\label{eq:hydro_end}
\end{align}
\noindent
The set of equations resulting from the transformation are the same as the original equations of hydrodynamics in a non-expanding system.
This formulation 
allows the extension of the hydrodynamics solver to an expanding frame system without any significant changes to the original solver.

\subsection{Hydrodynamics Solver}
\label{sec:hydro}

Without the gravitational source terms, Equations \ref{eq:hydro_start}-\ref{eq:hydro_end} correspond to the conserved form of the Euler equations. A detailed description of the methodology used for solving the gravity-free fluid dynamics can be found in \cite{schneider2015a}. The hydrodynamics solver is a Godunov-based method for which an approximation to the cell averaged values of the conserved quantities $\mathbf{U}=[\rho, \rho \mathbf{v}, \rho E, \rho e]$ are evolved using a numerical discretization of the Euler equations, given by 
\begin{equation}
\frac{\mathbf{U}_{i}^{n+1}-\mathbf{U}_{i}^{n}}{\Delta t}+\frac{\mathbf{F}_{i+1 / 2}^{n+1 / 2}-\mathbf{F}_{i-1 / 2}^{n+1 / 2}}{\Delta x}=0
\label{eq:hydro_fluxes}
\end{equation}
\noindent
where $\mathbf{U}_{i}^{n}$ denotes the average value of the conserved quantities for cell $i$ at  time-step $n$. The change of the conserved quantities in cell $i$ is given by the time centered fluxes across the cell interfaces $\mathbf{F}_{i \pm 1 / 2}^{n+1 / 2}$. The flux components $\mathbf{F}=[\rho \mathbf{v}, \rho \mathbf{v} \mathbf{v}, (\rho  E + p)\mathbf{v}, \rho  e \mathbf{v} ]$ are computed by solving the Riemann problem at the cell interfaces using the reconstructed values of the conserved quantities obtained via a Piecewise Parabolic Method \citep[PPM;][]{Colella+1984}. The PPM scheme is third-order accurate in space and second-order accurate in time.

\subsection{Dual Energy Implementation}
\label{sec:dual_energy}

Owing to the supersonic flows from structure formation
and adiabatic cooling of gas from universal expansion,
regions where the 
gas kinetic energy is much larger than the internal energy
are common in cosmological simulations.
Under these conditions,
calculation of the internal energy $E- v^2/2$ can be affected by numerical errors.
These errors can be ameliorated by using a ``dual energy formalism'' \citep{bryan1995a},
where the internal energy is evolved separately via Equation \ref{eq:internal_energy_1}, or the corresponding simplified Equation \ref{eq:hydro_end}, and substituted for the internal
energy computed from the total and kinetic energies when appropriate.
The two terms on the right side of Equation \ref{eq:hydro_end} correspond to
advection and compression terms, respectively.
To reconcile the total internal energy $E$ with the separately tracked internal energy $e$,
at the end of each time step a condition is applied on a cell to cell basis to select which value of the internal energy to employ. We adopt a condition similar to that used in \Enzo \citep{bryan2014a},
where the selection is based on the fraction of the internal energy in a given cell relative
to the maximum of the total energy in a the neighborhood of the cell. Mathematically,
this condition is given by
\begin{equation}
\label{eq:dual_energy_condition}
e_{i}=\left\{\begin{array}{ll}{e_i,} & {\rho_{i}\left(E-\mathbf{v}^{2} / 2\right)_{i} / \max(\rho E)_{i} <\eta} \\ {\left(E-\mathbf{v}^{2} / 2\right)_{i},} & {\text { otherwise }} \end{array}\right.,
\end{equation}
\noindent
where $\max(\rho E)_i$ is the maximum total specific energy in the local and adjacent cells.
In one dimension,  $\max(\rho E)_i = \max\left[( \rho E)_{i-1}, (\rho E)_{i}, (\rho E)_{i+1}\right]$. 
At the end of every time step, after applying Equation \ref{eq:dual_energy_condition}, the total specific energy $E$ is synchronized with the selected internal energy by setting $E = e + v^2/2$.

The value of $\eta$ should be chosen carefully, as setting $\eta$ too low will allow spurious heating owing to numerical errors introduced in the total energy evolution.
If $\eta$ is set to high then shock heating in regions where the gas flows converge could be suppressed since the advected internal energy $e$ will be preferentially selected over the conserved internal energy $E-v^2/2$,
and Equations \ref{eq:internal_energy_1} and \ref{eq:hydro_end} do not capture shock heating.
To estimate an appropriate value for $\eta$ in cosmological simulations, we developed a test to
evaluate how the dual energy condition affects the average cosmic gas
temperature, as described below in \S \ref{sec:avrg_temp_comparison}. 
Based on the results of this test, we set $\eta=0.035$.

Another approach on the selection criteria for the internal energy is presented in \cite{Teyssier2015}.
Here, the conserved internal energy $\rho(E-v^2/2)$ is compared to an estimate of the
numerical truncation error
\begin{equation}
e_{\text{trunc}} \simeq \frac{1}{2} \rho (\Delta v)^2,
\label{eq:truncation_error}
\end{equation}
\noindent
where $\Delta v$ corresponds to the difference of the velocities in the neighboring cells.
The selection condition for this scheme is given by
\begin{equation}
\label{eq:dual_energy_condition_ramses}
e_{i}=\left\{\begin{array}{ll} {\left(E-\mathbf{v}^{2} / 2\right)_{i}}  & {,\rho_{i}\left(E-\mathbf{v}^{2} / 2\right)_{i} > \beta e_{\text{trunc}}} \\ {e_i} & {,\text { otherwise }} \end{array}\right.,
\end{equation}
\noindent
where $\beta$ is a numerical parameter with suggested value $\beta=0.5$.
We also evaluated this dual energy condition using 
the average cosmic temperature test described in \S \ref{sec:avrg_temp_comparison}.
As we discuss below, we found that Equation \ref{eq:dual_energy_condition_ramses} 
can be overly restrictive by predominately selecting the advected internal energy $e$ over the conserved internal energy $E-v^2/2$, thereby
suppressing shock heating inside collapsed halos and significantly lowering the average cosmic temperature.

\subsection{Gravity}
\label{sec:gravity}

The gravitational acceleration vector $\mathbf{g}$ is computed by differentiating the gravitational potential $\phi$. The potential is obtained from the solution of the Poisson equation. In the comoving coordinates,
the Poisson equation is written as
\begin{equation}
\label{eq:poisson}
\nabla ^ 2 \phi = \frac{4 \pi G }{a} ( \rho  - \bar{\rho} ),
\end{equation}
\noindent
where $G$ is the gravitational constant, $\rho = \rho_{\mathrm{DM}} + \rho_b $
is the total dark plus baryonic matter density,
and $\bar{\rho}$ is the average value of the total density over the entire box.

Integration of Equation \ref{eq:poisson} can be directly performed in Fourier space.
In $k$-space, the Poisson equation simplifies to
\begin{equation}
 \hat{\phi}(\mathbf{k}) = G(\mathbf{k}) \hat{\rho}(\mathbf{k}),
\label{eq:poisson_fourier}
\end{equation}
\noindent
where $G(\mathbf{k})$ is the Greens function, which for a second-order centered two point finite difference discretization corresponds to \citep{Hockney+1988}

\begin{equation}
\label{eq:greens_function}
G(\mathbf{k})=-\frac{  \Delta x_{h} ^2}{\sin^{2} \left( k_{x} \Delta x_{h}\right)} -\frac{ \Delta y_{h} ^2}{\sin^{2} \left( k_{y} \Delta y_{h}\right)} - \frac{ \Delta z_{h} ^2}{\sin^{2} \left(k_{z} \Delta z_{h}\right)}.
\end{equation}
\noindent
Here $\Delta x_h = \Delta x /2$, $\Delta y_h = \Delta y /2$, and $\Delta z_h = \Delta z /2$, where
$\Delta x$, $\Delta y$, and $\Delta z$ are the grid cell dimensions.
To compute the three-dimensional fast Fourier transforms (FFTs)
we use \emph{PFFT} \citep{PFFT}, a publicly available
library for performing FFTs with a box domain decomposition.   

From the potential $\phi$ we compute the gravitational acceleration vector $\mathbf{g} = - \nabla \phi$. The derivatives along each direction are obtained using a fourth-order centered four-point finite difference
approximation. In one dimension, the derivative is given by

\begin{equation}
\label{eq:pot_gradient}
\frac{\partial \phi_{i}}{\partial x}  = \frac{1}{12 \Delta x} \left( \phi_{i-2} - 8 \phi_{i-1} + 8 \phi_{i+1} - \phi_{i+2}  \right).
\end{equation}
\noindent

The terms corresponding to the gravitational sources, $\tilde{\rho_b} \mathbf{\tilde{g}} $ and $ \tilde{\rho_b} \mathbf{\tilde{v}} \cdot \mathbf{\tilde{g}} $ in Equations \ref{eq:hydro_momentum} and \ref{eq:hydro_energy},  are added to the momentum and total energy after the conserved variables have been updated by the hydro solver 
(i.e., after Equation \ref{eq:hydro_fluxes} has been solved).
The numerical implementation for the coupling of the momentum and
energy with gravity is given by

\begin{align}
\label{eq:gravity_coupling}
\begin{split}
  (\tilde{\rho_b} \tilde{\mathbf{v}})_{i}^{n+1}    & =(\tilde{\rho_b} \tilde{\mathbf{v}})_{i}^{n+1*} \\ 
  & \,\, +\frac{1}{2}\Delta t^n  \left( \tilde{\rho}_{b,i}^{n}+\tilde{\rho}_{b,i}^{n+1} \right) \tilde{\mathbf{g}}_{i}^{n+1/2},
\end{split}
\\
\begin{split}
    (\tilde{\rho_b} \tilde{ E})_{i}^{n+1}&=(\tilde{\rho_b}  \tilde{E})_{i}^{n+1*} \\ 
    &  \,\, +\frac{1}{2}\Delta t^n \left[ (\tilde{\rho_b} \tilde{\mathbf{v}})_i^{n} +  (\tilde{\rho_b} \tilde{\mathbf{v}})_i^{n+1*} \right] \tilde{\mathbf{g}}_{i}^{n+1/2}.
\end{split}
\end{align}
\noindent
Here the superscript $n+1*$ refers to the value of the conserved quantity after the hydrodynamics solver update and the time centered value of the gravitational field  $\tilde{\mathbf{g}}_{i}^{n+1/2} = - \nabla \tilde{ \phi_i }^{n+1/2}$. 
The potential $\tilde{ \phi_i }^{n+1/2}$  is obtained by extrapolation from  $\phi_i^{n}$ and $\phi_i^{n-1}$ using
\begin{equation}
\label{eq:pot_extrapolation}
\tilde{ \phi_i }^{n+1/2} = \frac{(a^n)^2}{H_0^2} \left[ \phi_i^n  + \frac{\Delta t^n}{ 2 \Delta t ^{n-1}} ( \phi_i^n - \phi_i^{n-1} )  \right].
\end{equation}

\subsection{Dark Matter }
\label{sec:dark_matter_particles}

We represent the cold dark matter as a system of discrete point-mass particles moving under the influence of
gravity.
Each dark matter particle is described by its mass $m_i$, comoving position $\mathbf{x}_i$,
and peculiar velocity $\mathbf{v}_i$. The evolution of the particle
trajectories in a comoving frame is described by
\begin{align}
\frac{d \mathbf{x}_{i}}{d t} &= \frac{1}{a} \mathbf{v}_{i} \\ 
\frac{d\left(a \mathbf{v}_{i}\right)}{d t} & =\mathbf{g}_{i},
\end{align}
\noindent
where $\mathbf{g}_i$ is the acceleration vector owing
to the gravitational field evaluated at the particle position $\mathbf{x}_i$.

To solve Equation \ref{eq:poisson}, we compute the contribution of the dark matter particles to the density field
by interpolating onto the same grid used to evolve the hydrodynamical quantities.
The dark matter density $\rho_{\mathrm{DM}}$ is calculated via a cloud-in-cell scheme \citep{Hockney+1988}, for which each particle is represented as a cube having the same size as one grid cell $\Delta x$ and uniform density $m_i/ \Delta x^3$.
The mass of the particle is distributed among the grid cells that intersect its volume such that the fraction of the particle mass $\delta m_i$ deposited on a cell is equal to the fraction of its intersected volume.

In one dimension, the mass contribution of a particle to a cell at position $x_c$ is given by
\begin{equation}
\delta m_{i,c} = m_i \left\{\begin{array}{ll}{1-|x_i - x_c| / \Delta x,} & {|x_i - x_c|<\Delta x} \\ {0,} & {\text { otherwise }}\end{array}\right..
\end{equation}
\noindent
The gravitational acceleration $\mathbf{g}_i$ evaluated at a particle position $\mathbf{x}_i$ must be computed in a  manner consistent with the particle density interpolation. For the cloud-in-cell scheme, to avoid self-forces on the
particles each component of $\mathbf{g}$ should be interpolated with the same weights used during
the density assignment calculation.

To integrate the particle trajectories we use the kick-drift-kick (KDK) method \citep{Miniati+2007},
consisting of three steps to update the particles position and velocity from time-step $n$ to time-step $n+1$.
The sequence of variable updates is
\begin{align}
\label{eq:particles_advance_1}
\mathbf{v}_i^{n+1/2} &= \frac{1}{a^{n+1/2}} \left(  a^n \mathbf{v}_i^n  + \frac{\Delta t^n}{2} \mathbf{g}_i^n \right), \\
\label{eq:particles_advance_2}
\mathbf{x}_i^{n+1} &= \mathbf{x}_i + \frac{\Delta t^n}{ a^{n+1/2}} \mathbf{v}_i^{n+1/2}, \\
\label{eq:particles_advance_3}
\mathbf{v}_i^{n+1} &= \frac{1}{a^{n+1}} \left(  a^{n+1/2} \mathbf{v}_i^{n+1/2}  + \frac{\Delta t^n}{2} \mathbf{g}_i^{n+1} \right). 
\end{align}
\noindent
The KDK scheme allows for variable timesteps, as required by cosmological simulations owing
to the variation in gas and particles velocities as the simulation advances.
This sympletic scheme conserves an integral of motion on average,
preventing an accumulation of errors and maintaining the phase space
trajectory of the particles.

\subsection{Chemistry and Radiative Cooling}
\label{sec:cooling}

We integrated \Cholla with the GRACKLE chemistry and cooling library \citep{smith2017a} to
solve a non-equilibrium chemical network.
Currently, our method only tracks the atomic chemical species and metals,
but it could be extended to include, e.g., molecular hydrogen and deuterium.

The chemical species (\ion{H}{1}, \ion{H}{2}, \ion{He}{1}, \ion{He}{2}, \ion{He}{3}, electrons $\text{e}^-$,
and metals $\text{Z}$)
are advected as scalar fields alongside the gas conserved variables via Equation \ref{eq:hydro_fluxes}.
For details about the implementation of GRACKLE, we refer the reader to \citet{smith2017a}.
During every time step, GRACKLE updates the ionization fractions and computes the net heating and cooling by sub-cycling the rate equations within one hydrodynamic step.
The sub-cycling updates the chemical and thermal states of the gas on timescales smaller than the dynamical timescales.
For the atomic H and He chemical network, GRACKLE directly computes the heating and cooling rates
accounting for collisional excitation and ionization, recombination, free-free emission, Compton scattering from the cosmic microwave background, and photoheating from a metagalactic UV background.
GRACKLE accounts for metals by using precomputed tables for the metallic cooling and heating rates.

The GRACKLE update routine is applied at the end of each time step,
after the gas conserved variables have been updated by the hydro solver and additional gravitational source terms.
This routine updates the ionization fraction of the chemical elements, and also adds
the net cooling and heating to the internal energy by setting
\begin{align}
(\rho e)_i^{n+1}  & \rightarrow  (\rho e)_i^{n+1}  + a  \Delta t^n ( \Gamma - \Lambda )_i^n
\label{eq:e_update}
\end{align}
\noindent
Finally,  the total energy is updated to reflect the change in the internal energy due to the net cooling as
\begin{equation}
E_i^{n+1} = \frac{1}{2}(v_i^{n+1})^2 + e_i^{n+1}. 
\end{equation}

\subsection{UVB Ionization and Heating}
\label{sec:uvb_heating}

The non-equilibrium GRACKLE solver accounts for the ionization of the primordial chemical species owing to a uniform time-dependent UV background 
by loading tables of the redshift dependent photoionization and photoheating rates for \ion{H}{1}, \ion{He}{1},
and \ion{He}{2}. We compute the photoionization rates from a given redshift dependent spectrum as
\begin{equation}
\Gamma_{\gamma i}(z) = \int_{\nu_{i}}^{\infty} \frac{4 \pi J(\nu, z)}{h \nu} \sigma_{i}(\nu) d \nu,
\end{equation}   
\noindent
where $J(\nu, z)$ is the intensity of the UV background at frequency $\nu$ (in $\mathrm{erg} \, \mathrm{s}^{-1} \, \mathrm{cm}^{-2}\, \mathrm{sr}^{-1}\, \mathrm{Hz}^{-1}$), and $\nu_i$ and $\sigma_i(\nu)$ are the threshold frequency and cross-section for photoionization of the species $i$, taken from \cite{Osterbrock1989}. Analogously,
the photoheating rates are computed as
\begin{equation}
\epsilon_{i}(z) = \int_{\nu_{i}}^{\infty} \frac{4 \pi J(\nu, z)}{h \nu} (h\nu - h\nu_i) \sigma_{i}(\nu) d \nu.
\end{equation}   
\noindent

If metal line cooling is included, the contributions of metals to the heating and cooling rates are
accounted by GRACKLE by loading precomputed density, temperature and redshift dependent lookup tables 
that were obtained by providing the UVB spectrum to the CLOUDY \citep{Cloudy2017} photoionization code (version 17.02). 
The tables for metallic heating and cooling rates were generated for solar metallicity under the assumption of ionization equilibrium and subtracting the contributions of primordial heating and cooling as described in \cite{smith2017a}. The resulting tables are organized into a Hierarchical Data Format (version 5) file readable by GRACKLE.

\subsection{ Time Step Calculation }
\label{sec:time_step}

The simulation time step $\Delta t$ is computed with constraints from the signal speed of the gas,
the motion of the dark matter particles, and the expansion of the universe.
For the gas, the time step is constrained by the gas velocities $\mathbf{v}$ and the sound speed $c_s$
as

\begin{equation}
\Delta t_{\mathrm{gas}}=\alpha_{\mathrm{gas}} \min \left(\frac{a \Delta x}{|v_x|+c_s}, \frac{ a  \Delta y}{  |v_y|+c_s}, \frac{ a \Delta z}{|v_z|+c_s}\right),
\end{equation}
\noindent
where $\alpha_{\mathrm{gas}}$ is the CFL factor specified by the user ( $\alpha_{\mathrm{gas}}=0.3$ by default) and $|v| + c_s$ is evaluated  over the entire grid for each direction to find  the minimum value of $\Delta t_{\mathrm{gas}}$.
For the particles, the time step is limited to avoid any displacement larger than the cell size in each direction using
\begin{equation}
\Delta t_{\mathrm{DM}}=\alpha_{\mathrm{DM}} \min \left(\frac{a \Delta x}{|v_x|}, \frac{ a \Delta y}{|v_y|}, \frac{ a \Delta z}{|v_z|}\right),
\end{equation}
\noindent
where $\alpha_{\mathrm{DM}}$ is analogous to the CFL factor ($\alpha_{\mathrm{DM}}=0.3$ by default) and ${|v|}$ is evaluated over all the particles for each direction.
The time step is also limited by the expansion of the universe by choosing $\Delta t_{\mathrm{exp}}$ such that the fractional
change in the scale factor does not exceed 1\% ($\Delta a_{\mathrm{exp}} = 0.01 a$).
The actual time step is selected by taking the smallest value
\begin{equation}
\Delta t= \min \left( \Delta t_{\mathrm{gas}}, \Delta t_{\mathrm{DM}}, \Delta t_{\mathrm{exp}}\right),
\label{eq:delta_t}
\end{equation}
guaranteeing that all three limiting conditions described above are satisfied.

The time step $\Delta t$ (Eq. \ref{eq:delta_t}) is applied globally to update all the cells and particles in the box. When running high resolution simulations, it is possible to have a situation in which for a single cell $\Delta t_{\mathrm{gas},i}$ is extremely small compared to the all the other cells, resulting in small values for $\Delta t$ which significantly slow the entire simulation. To avoid this situation, when a cell satisfies the condition $\Delta t_{\mathrm{gas},i} < \Delta t_{\mathrm{DM}} /50  $ then the conserved quantities of that cell (density, momentum, energy, and internal energy) are replaced with the conserved quantities averaged over the six closest neighboring cells, resulting in a larger $\Delta t_{\mathrm{gas},i}$ for such cell and avoiding extremely small steps. We keep track of this occurrences during the full run, and for the high resolution simulations presented in this work this situation happens less than a dozen times per simulation, ensuring that the dynamics of the gas are not significantly affected by these small interventions.

\subsection{Algorithm Implementation}
\label{sec:algorithm}

The complete method to evolve the gas and the dark matter particles from time-step $n$ to time-step $n+1$ 
can be summarized by the following algorithm:\\

\textbf{Initialization:}

\begin{enumerate}

\item Load initial conditions for the gas conserved variables and the particle positions and velocities.

\item Obtain the dark matter density $\rho_{\mathrm{DM}}$ by interpolating the particle masses onto the grid via the Cloud-In-Cell method described in \S\ref{sec:dark_matter_particles}.

\item Compute the gravitational potential $\phi$ by solving the Poisson equation (Eqn.~\ref{eq:poisson}), using the dark matter density $\rho_{\mathrm{DM}}$ and the gas density $\rho_b$ as the sources.

\item Calculate the gravitational field $\mathbf{g}=-\nabla \phi$ at the centers of the grid cells using a fourth-order finite difference scheme (Eqn.~\ref{eq:pot_gradient}) and interpolate the acceleration vector evaluated at the particles positions $\mathbf{g_i}=\mathbf{g}(\mathbf{x}_i)$.

\end{enumerate}

\textbf{Time Step Update:}

\begin{enumerate}

    \item Compute the current time step $\Delta t^n$. 
    
    \item Obtain the gravitational potential at $t^{n+1/2}$ by extrapolation using $\phi^{n}$ and $\phi^{n-1}$, Equation \ref{eq:pot_extrapolation}.
    
    \item Advance the gas conserved quantities by $\Delta t^n$ using the intercell fluxes $\mathbf{F}^{n+1/2}$, Equation \ref{eq:hydro_fluxes}.
    
    \item Add the gravitational sources to the gas momentum and energy, Equation \ref{eq:gravity_coupling}. 
    
    \item Call GRACKLE to update the ionization states of the chemical network and add the net cooling and heating to the internal energy, Equation \ref{eq:e_update}. 

    \item Advance the particle velocities by $\frac{1}{2} \Delta t^n$ and use the updated velocities $\mathbf{v}_i^{n+1/2}$ to advance the particle positions by $\Delta t^n$, Equations \ref{eq:particles_advance_1} and \ref{eq:particles_advance_2}.
    
    \item Obtain the dark matter density $\rho_{\mathrm{DM}}^{n+1}$ via the CIC method.
    
    \item Compute the gravitational potential $\phi^{n+1}$ by solving Equation \ref{eq:poisson} with $\rho_{\mathrm{DM}}^{n+1}$ and $\rho_{b}^{n+1}$ as sources.
    
    \item Obtain the gravitational field $\mathbf{g}^{n+1}$ at the cell centers  and $\mathbf{g}_i^{n+1}$ evaluated at the particle positions.
    
    \item Advance the particle velocities by $\frac{1}{2} \Delta t^n$ resulting in $\mathbf{v}_i^{n+1}$, Equation \ref{eq:particles_advance_3}.
   
\end{enumerate}

Currently, the extensions included into \Cholla for cosmological simulations (the FFT based Poisson solver, the dark matter particles integrator and the chemical network solver) all are implemented to run in the host CPUs, while the hydrodynamics solver including the advection of the ionization states of H and He run in the GPUs. At the time of submission of this work, an entirely GPU based distributed FFT solver has been recently integrated into \Cholla and development to transfer both the particle integrator and the H+He network solver to the GPUs is ongoing. Potentially, the GPU implementation of a H+He network solver analogous to GRACKLE could result in a significant performance increase since currently the GRACKLE call to update the chemical network is the slowest step in our implementation.

\section{Validation}
\label{sec:validation}

To test the extensions of the \Cholla code for cosmological simulations, we
present below a set of validation exercises including comparisons with 
other publicly available Eulerian codes. In \S \ref{sec:zeldovich} we
present the standard \citet{zeldovich1970a} test. We then compare 
in \S \ref{sec:dark_matter_simulations} the matter
power spectra of N-body cosmological simulations performed with 
\Cholla to results from the \Nyx \citep{almgren2013a}, \Ramses \citep{teyssier2002a},
and \Enzo \citep{bryan2014a} codes using the same initial conditions,
and find sub-percent-level agreement at all spatial scales when
simulated with the same resolution.
We extend these tests to adiabatic hydrodynamical cosmological
simulations in \S \ref{sec:adiabatic_hydro_sim}, where we find agreement
within a few percent. To test the dual energy formalism in cosmological simulations,
we describe a new test that computes the mean gas temperature with redshift
(\S \ref{sec:avrg_temp_comparison}), and show that our choice of
dual energy paramterization and parameter values recovers model
expectations. We validate our cosmological hydrodynamical
simulations including cooling, chemistry, and heating against \Enzo
simulations using the same physical prescription, and find good agreement.

\subsection{Zel'Dovich Pancake}
\label{sec:zeldovich}

\begin{figure}
\includegraphics[width=0.47\textwidth]{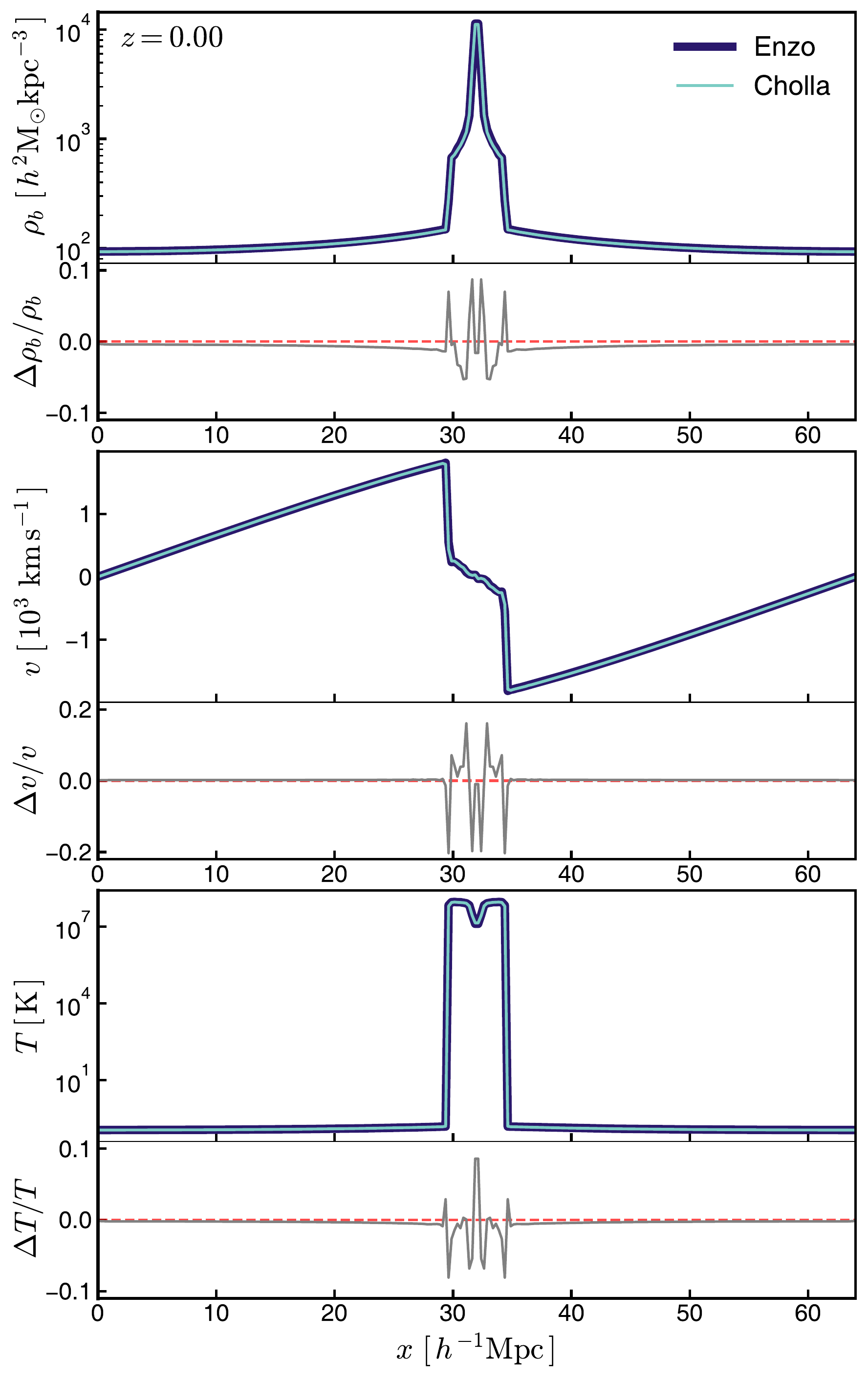}
\caption{Zel'Dovich pancake test at $z=0$ for a one dimensional 64 $h^{-1}$Mpc box discretized on a 256 uniform grid for a $h=0.5$ and $\Omega_M=1$ universe. Shown are the solutions computed using \Enzo (purple) and \Cholla (blue). The top, middle, and bottom main panels correspond to the density ($\rho_b$), velocity ($v$), and temperature ($T$), respectively, the fractional differences for each quantity are shown at the bottom of each panel (gray lines). The \Cholla simulations resolve the central shock and overdensity, and the results are in excellent agreement with the \Enzo simulation. Small differences $ < 10\%$ for $\rho_b$ and $T$ are located at the sharp features of the shock, and differences $\sim 20 \%$ for $v$ at the regions where $v \sim 0$ and at the fronts of the shock.     }
\label{fig:zeldovich}
\end{figure}

The \citet{zeldovich1970a}  pancake problem encompasses several of the basic components of a cosmological hydrodynamical simulation including gas dynamics, self-gravity, and an expanding frame.
For this test, the evolution of a single one dimensional sinusoidal perturbation is followed to provide a useful representation of the gas evolution in a three-dimensional simulation by solving the gravitational collapse of a single mode. The initial conditions for the density, velocity and temperature on a Lagrangian frame are set as
\begin{align}
\rho_b\left(x_{l}\right) &=\rho_{0}\left[1-\frac{1+z_{s}}{1+z} \cos \left(k x_{l}\right)\right]^{-1} \\
v\left(x_{l}\right)&=-H_{0} \frac{1+z_{s}}{(1+z)^{1 / 2}} \frac{\sin \left(k x_{l}\right)}{k} \\
T\left(x_{l}\right)&=T_{\text {0}}\left[\frac{\rho_b\left(x_{l}\right)}{\bar{\rho_b}}\right]^{2 / 3},
\end{align}
\noindent
where $z_s$ is the value of the redshift at which the gravitational collapse results in the formation of a shock located  at the center of the overdensity, $z$ is the initial redshift, $\lambda$ is the wavelength of the perturbation , $k=2\pi/\lambda$ is the corresponding wavenumber, and $x_l$ is the position of the Lagrangian mass coordinate.
The conversion of the positions to the Eulerian coordinates $x$ is given by
\begin{equation}
x=x_{l}-\frac{1+z_{s}}{1+z} \frac{\sin \left(k x_{l}\right)}{k}.
\end{equation}

\begin{figure*}[t!]
\includegraphics[width=\textwidth]{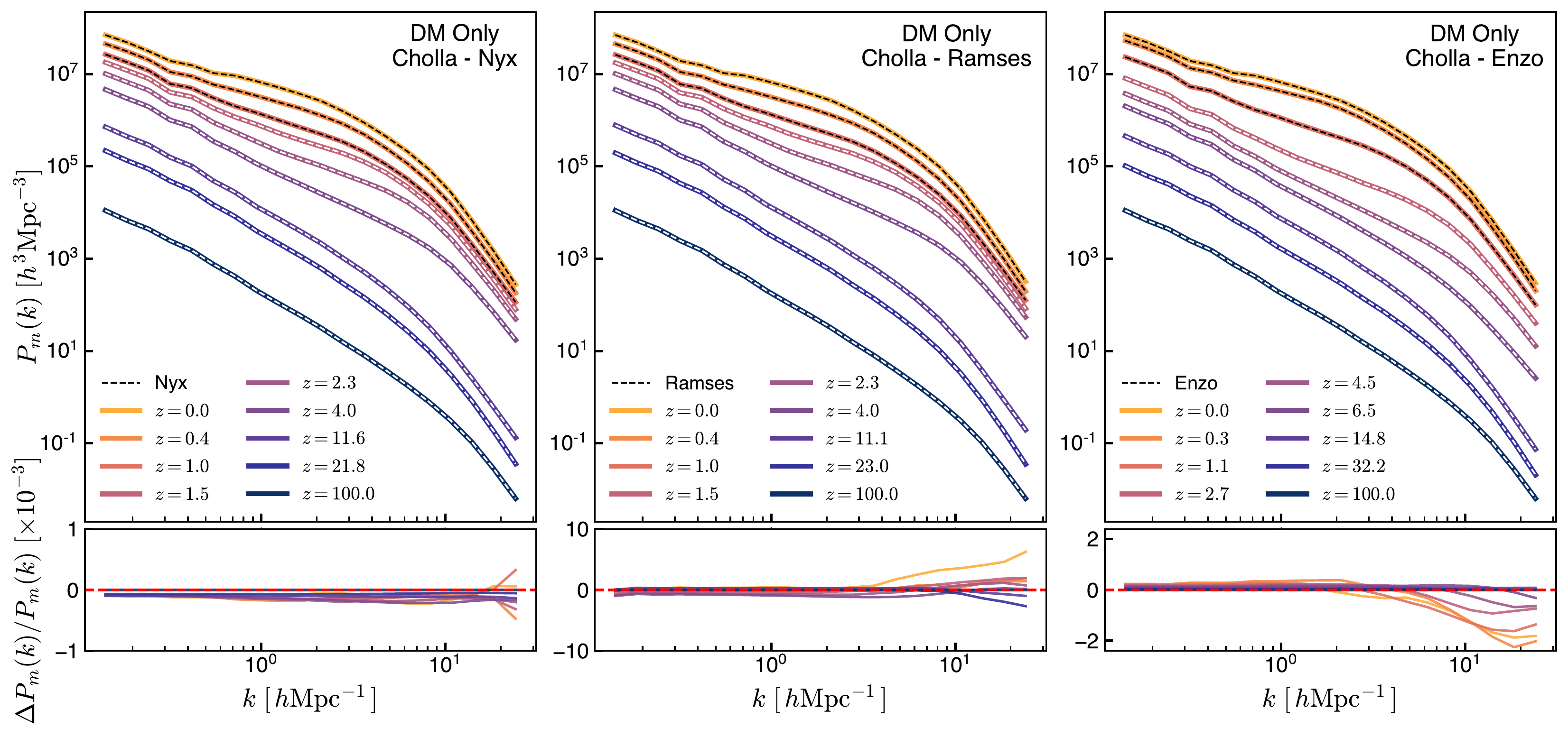}
\caption{Matter power spectrum $P_m(k)$ from dark matter-only simulations. Shown are $P_m(k)$ from simulations performed with \Cholla (colored lines) compared with analogous simulations computed with \Nyx, \Ramses, and \Enzo (left, center, and right panels; dashed lines).
The bottom panels show the fractional difference between the \Cholla simulation and each corresponding code for comparison. For each of the comparison codes, the fractional differences relative to the \Cholla results are $<0.1$\% compared with \Nyx (bottom left), $<1$\% compared with \Ramses (bottom center), and $<0.3$\% compared with \Enzo (bottom right).}
\label{fig:power_spectrum_dm}
\end{figure*}
\noindent
For this test we replicate the problem presented in \cite{bryan2014a}, a one dimensional simulation on an $L=64h^{-1}$Mpc box with cosmological parameters $h=0.5$ and $\Omega_M=1$.
The initial background density and temperature ( $\rho_0$ and $T_0$ ) are set equal to the critical density $\rho_c =3 H_0^2 / 8 \pi G$ and $100\,\mathrm{K}$ respectively.
The redshift at which the shock develops is set as $z_s=1$ and the wavelength of the perturbation is set as $\lambda = L = 64h^{-1}$Mpc. The simulation is initialized at $z=20$ and runs on an  $N=256$ cell uniform grid.

Comparing the evolution of the Zel'dovich pancake problem solved using \Cholla and \Enzon, we find
the results from the \Cholla simulation closely resemble the \Enzo results.
Both simulations develop a central shock at $z=1$ and the central overdensity grows at the same rate.
Figure \ref{fig:zeldovich} compares the density $\rho_b$, velocity $v$ and temperature $T$ fields at $z=0$ in
the simulations solved with \Enzo (purple lines) and \Cholla (blue lines), additionally, the fractional differences are shown at the bottom of each panel (gray lines). We measure small differences $< 10 \%$ for the density and the temperature only at the sharp features of the shock. For the velocity the differences are $\sim 20\%$ at the discontinuities located at the front of the shock, and at the regions where $v \sim 0$, the other regions contained by the shock show small differences $< 10 \%$. The regions outside the shock result in differences $< 1\%$ for all the fields. The small differences demonstrate an excellent agreement between the codes.

We note that we also performed the Zel'Dovich pancake test by
applying Equation \ref{eq:dual_energy_condition_ramses} for the internal energy selection in the dual energy scheme.
This condition causes the code to select the advected internal energy $e$ instead of the
conserved internal energy $E-v^2/2$ during the entire simulation.
Since Equation \ref{eq:internal_energy_1} does not captures shock heating, the 
central shock at $z=1$ is suppressed and the temperature of the central region only gradually increases
owing to the gas compression.
This behavior results in a significantly different distribution for the density, velocity, and temperature in the central region. We discuss further ramifications of the dual energy condition in \S \ref{sec:avrg_temp_comparison}.

\subsection{ N-body Cosmological Simulations}
\label{sec:dark_matter_simulations}

To validate the results produced by \Cholla in a realistic cosmological setting, first we compare \Cholla dark matter-only simulations with calculations using several other well-established codes.
In this test, the simulation domain consists of  an $L=50(h^{-1}\mathrm{Mpc})^3$ box. The standard
cosmological parameters are set to $\Omega_M=0.3111$, $\Omega_\Lambda=0.6889$, $h= 0.6766$, and $\sigma_8= 0.8102$. 
Initial conditions were generated at $z=100$ on a uniform resolution grid using the MUSIC software \citep{MUSIC}. For this test, we evolve $256^3$ particles on an $N=256^3$ cell uniform grid, with the particle mass resolution equal to $m_p=6.4345 \times 10^8 h^{-1} \text{M}_\odot$. 

For our validation test, we measure the matter power spectrum $P_m(k)$ evolved from identical initial conditions using
\emph{Cholla}, \Nyx \citep{almgren2013a}, \Ramses \citep{teyssier2002a}, and \Enzo \citep{bryan2014a}.
To measure the matter power spectrum for each simulation, we first compute the dark matter density field by interpolating the dark matter particles onto the $N=256^3$ uniform grid via the CIC method described in
\S \ref{sec:dark_matter_particles}.
The power spectrum is then computed in Fourier space by taking the FFT of the overdensity field.
The density field and power spectrum are computed identically for all the comparison simulations
to ensure that any power spectrum differences arise solely from differences in the evolved particle
distribution.

The results of our comparison are presented in Figure \ref{fig:power_spectrum_dm}.
Each panel in the upper row shows the matter power spectrum $P_m(k)$ for several redshifts as computed by \Cholla (colored lines),
along with an overlay of the results from other codes (dashed lines).
The left panel shows the comparison to the \Nyx simulation, the center panel corresponds to the \Ramses comparison, and the right panel shows the comparison to the \Enzo results.
The bottom row shows the fractional difference between the power spectrum of the simulation computed by \Cholla and
each comparison code. As the left lower panel shows, the power spectrum measured in the \Cholla and \Nyx simulations is in excellent agreement with fractional differences of $\approx0.05$\% at small scales, and even smaller differences of $\approx0.02$\% on larger scales.
The comparison with \Ramses also shows remarkable agreement with differences of 0.1\% at large scales, with the
largest differences of $\sim0.7\%$ occurring on small scales by $z=0$.
\Cholla and \Enzo also show excellent agreement,
with differences of $<0.1\%$ at large scales and $<0.3\%$ on small scales.   

We note that in the version of \Nyx used in this comparison, a second order scheme was employed
to compute the gravitational potential gradient,
\begin{equation}
\label{eq:pot_gradient_nyx}
\frac{\partial \phi_{i,j,k}}{\partial x}  = \frac{1}{2 \Delta x} \left(  \phi_{i+1,j,k} - \phi_{i-1,j,k}  \right),
\end{equation}
\noindent
instead of the fourth-order method described by Equation \ref{eq:pot_gradient}.
To have the closest possible comparison, for the \Cholla simulation used to compare to the \Nyx results we used Equation \ref{eq:pot_gradient_nyx} to compute the gravitational field. We note that the lower order scheme used to compute the gradient leads to significant differences on the power spectrum at small scales of about 15\% 
relative to the same simulation employing the higher order method.

Additionally, for the comparison with \Enzo we used the simpler kernel for the Greens function $G(\mathbf{k})=-k^{-2}$ in our solver instead of the kernel for the discretized Poisson equation given by Equation \ref{eq:greens_function}. 
The choice of the kernel results in substantial differences, 
changing the small-scale power spectrum by as much as $\approx28$\% relative to \Enzo
when \Cholla employs Equation \ref{eq:greens_function}.

\subsection{Adiabatic Cosmological Hydrodynamical Simulation }
\label{sec:adiabatic_hydro_sim}

\begin{figure}
\includegraphics[width=0.47\textwidth]{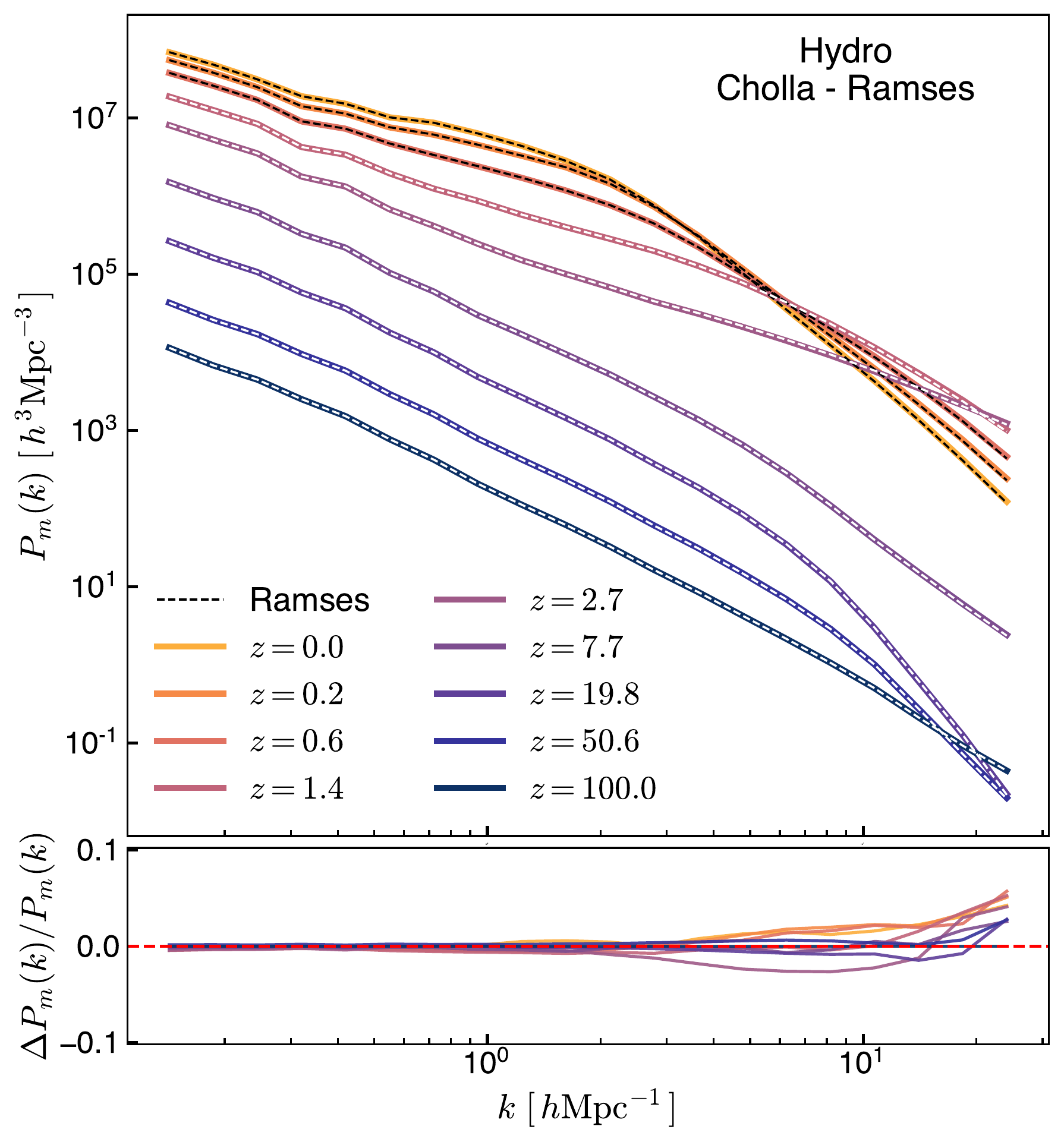}
\caption{Redshift-dependent gas density fluctuation power spectra for an adiabatic cosmological simulation.
Shown are the results from our \Cholla simulation (solid lines) compared with simulation evolved using \Ramses (dashed lines). The bottom panel shows the fractional difference between the \Cholla and \Ramses power spectra. The agreement is excellent on large scales, and for smaller scales the differences are $<7$\%.       }
\label{fig:power_hydro_ramses}
\end{figure}

In \S \ref{sec:zeldovich} we showed that \Cholla accurately solved the gas dynamics in a simplified one dimensional simulation. To test the evolution of the gas in a realistic cosmological evolution,
we compared the results of a \Cholla adiabatic hydrodynamical run to a calculation with \Ramses using
identical initial conditions.
The configuration for the simulation is the same as the one described in \S \ref{sec:dark_matter_simulations},
but with the addition of an $\Omega_b=0.0486 $ baryonic component. For the comparison we measure the gas density fluctuation power spectrum directly from the baryon density field in both simulations.
Figure \ref{fig:power_hydro_ramses} shows the results of the comparison, with
the power spectrum measured in the \Cholla simulation (colored lines) shown for several redshifts
along with the \Ramses simulation (dashed lines). The bottom panel shows the fractional difference between the \Cholla and \Ramses power spectrum measurement. On large scales the agreement is excellent ($\lesssim1$\%), and on
smaller scales there are some differences up to a maximum of $\approx7$\% at $z<1$.  

As described in \S \ref{sec:dual_energy}, the dual energy condition used by \Ramses (Equation \ref{eq:dual_energy_condition_ramses}) can suppress shock heating in regions where the gas is converging.
This choice can have a significant effect on gas falling into dark matter potential wells,
resulting in artificially low gas temperatures.
A detailed study of how the dual energy condition affects cosmological gas
properties is provided in \S \ref{sec:avrg_temp_comparison}, but for this power spectrum test
we used the \Ramses dual energy condition (Equation \ref{eq:dual_energy_condition_ramses}).
We note that the lower gas temperatures computed using the \Ramses dual energy condition
result in more power on small scales relative to calculations that use Equation \ref{eq:dual_energy_condition}.
The tests in \S \ref{sec:avrg_temp_comparison} illustrate why we instead use Equation \ref{eq:dual_energy_condition}
in our \Sim simulation suite.

\begin{figure}
\includegraphics[width=0.47\textwidth]{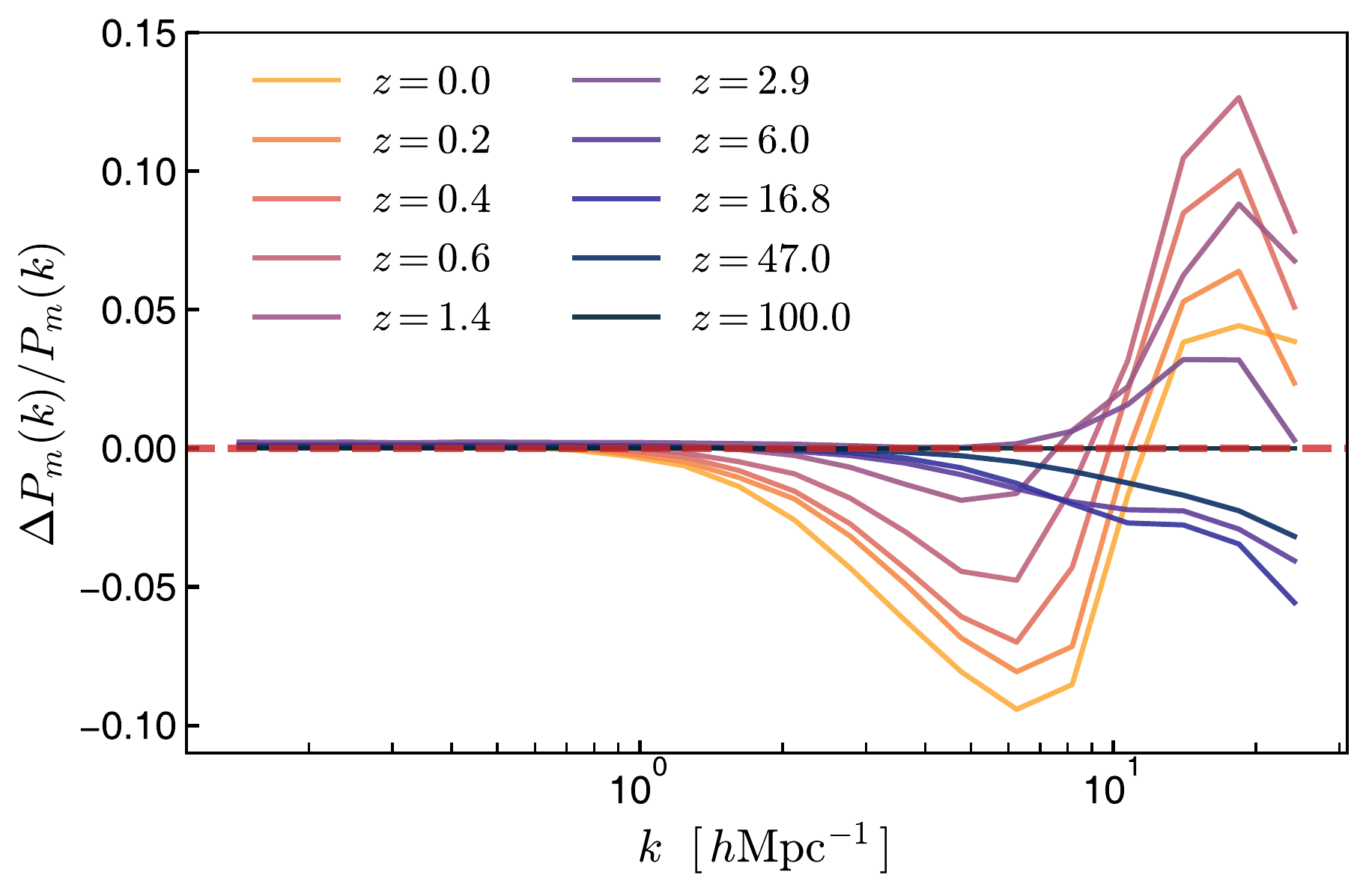}
\caption{Fractional differences in the gas density fluctuation power spectra for an adiabatic cosmological \Cholla simulation relative to an analogous \Enzo run, the differences are $< 1\%$ for large scales and $< 15\%$ for smaller scales. }    
\label{fig:power_hydro_enzo}
\end{figure}

Additionally we compared the power spectrum of the gas density fluctuations from a \Cholla adiabatic cosmological simulation using Equation \ref{eq:dual_energy_condition} for the dual energy formalism to an analogous \Enzo run that evolved the same initial conditions. Figure \ref{fig:power_hydro_enzo} shows the fractional differences in $P_m(k)$ resulting from the \Cholla simulation relative to the \Enzo run. As shown, the agreement is excellent in the large scales with differences $< 1\%$ and the smaller scales present differences $<15\%$. From the tests performed we note that the small scale $P_m(k)$ is highly sensible to the numerical implementation of the hydrodynamics solver and that we are not aware of a robust comparison of the gas $P_m(k)$ resulting from different codes. We argue that the small differences obtained do not represent an inaccurate evolution of the gas dynamics. 

\subsection{ Average Cosmic Temperature }
\label{sec:avrg_temp_comparison}

In \S \ref{sec:dual_energy} we discussed the dual energy formalism used when solving hydrodynamical cosmological simulations. We presented two different approaches for selecting between the advected internal energy $e$ or the conserved internal energy $E - v^2/2$, these two methods are given by Equations  
\ref{eq:dual_energy_condition} and \ref{eq:dual_energy_condition_ramses} employed by \Enzo  and  \Ramses respectively. 
To test which approach best captures the shock heating of the infalling gas onto the dark matter halos
when implemented in \Chollan,
we measure the mass weighted average gas temperature $\bar{T}$ in an adiabatic cosmological simulation
as described in \S \ref{sec:adiabatic_hydro_sim},
and compare the simulation results to an estimate of the expected gas temperature
computed from averaging the virial temperature of collapsed halos 
with the adiabatically-cooling IGM. 
We used the ROCKSTAR halo finder \citep{Behroozi+2013} to identify  dark matter halos in the \Cholla simulations, 
and then computed for each resolved halo a virial temperature as
\begin{equation}
    T_{vir} = \frac{m_p}{3 k_B} \frac{G M_{vir}}{R_{vir}}.
\end{equation}
\noindent
where $m_p$ is the proton mass, $k_B$ is the Boltzmann constant, and $M_{vir}$ and $R_{vir}$ are the virial mass and radius of the halo measured by ROCKSTAR.
To compute our reference estimate of the expected mean cosmic temperature in the simulation, 
we take the mass in the IGM as simply
total gas mass in the simulated box 
$M_{total}$ minus the mass in collapsed halos $M_{halos}=\sum M_{vir}$.
Then, assuming a uniform baryon fraction,
the fraction of gas mass present int the IGM is simply $(M_{total} - M_{halos})/M_{total}$, 
and from this we can compute the mass weighted average temperature as
\begin{equation}
    \bar{T}_{vir} =  \sum\limits_{halos} \frac{ M_{vir}}{M_{total}}  T_{vir}   + \left( \frac{M_{total} - M_{halos}}{M_{total}} \right) \left( \frac{a_0}{ a} \right)^2 T_0,
\label{eq:virial_temperature}
\end{equation}
\noindent
where the first term corresponds to the mass weighted virial temperature of the gas present in collapsed halos and the second term corresponds to the mass weighted temperature of the gas in the IGM.
The IGM temperature is taken to be the initial temperature $T_0$ scaled by the $a^{-2}$ factor
owing to the adiabatic expansion of the universe.

\begin{figure}[t!]
\includegraphics[width=0.47\textwidth]{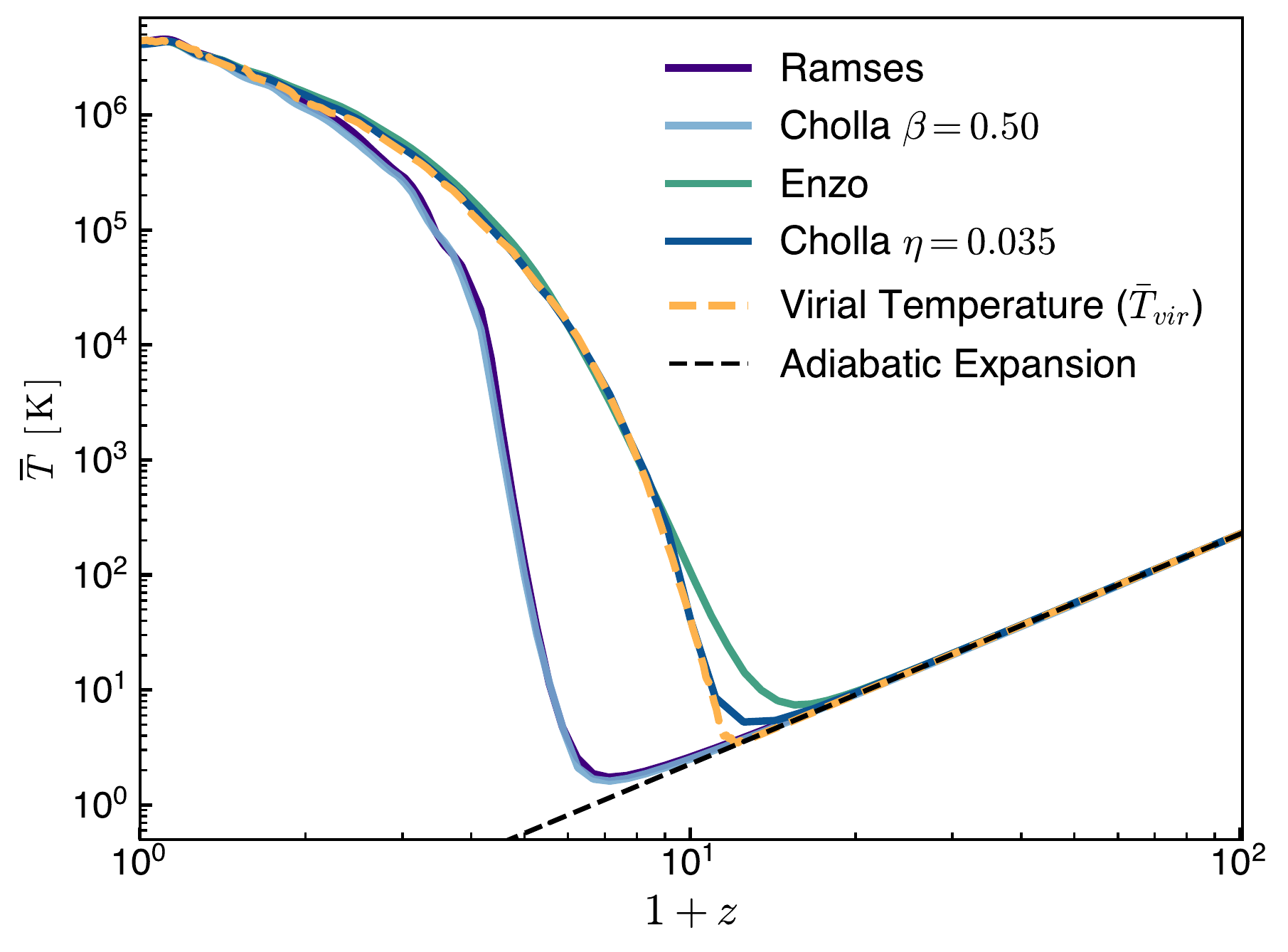}
\caption{Evolution of the mass weighted average cosmic temperature $\bar{T}$ as a function of redshift for an adiabatic cosmological simulation. Shown as solid lines are the simulation results using \Ramses (purple), \Enzo (green), and \Cholla simulations where Equations \ref{eq:dual_energy_condition} (dark blue) or \ref{eq:dual_energy_condition_ramses} (light blue) were used for the internal energy selection criteria in the dual energy formalism. The dashed lines show estimates of the temperature $T$ expected from the viral temperature of halos 
(yellow) and the $T\propto a^{-2}$ dependence owing to the adiabatic expansion of the universe (black).    }
\label{fig:avrg_temperature}
\end{figure}

\begin{figure}[t!]
\includegraphics[width=0.47\textwidth]{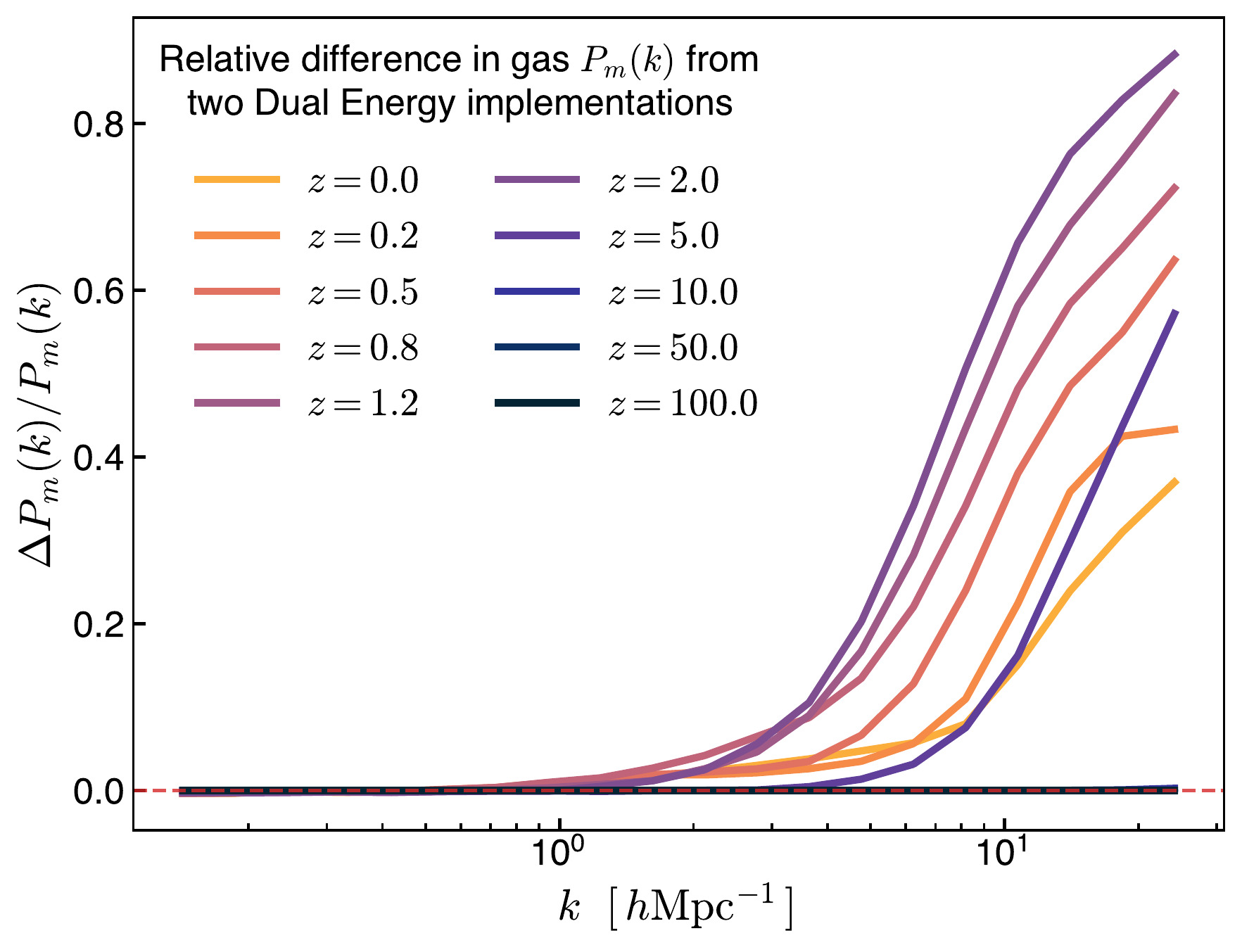}
\caption{Relative fractional difference of gas density fluctuation power spectra in adiabatic cosmological simulations using Equations \ref{eq:dual_energy_condition} or \ref{eq:dual_energy_condition_ramses} for the dual energy implementation. The condition set by Eqn.~\ref{eq:dual_energy_condition_ramses} maintains colder gas for $1 \lesssim z \lesssim 15$, resulting in larger power at small scales.
The differences range from 30\% to 80\% at $0 \lesssim z \lesssim 6$.}
\label{fig:ps_dual_energy_difference}
\end{figure}

\begin{figure*}[t!]
\includegraphics[width=\textwidth]{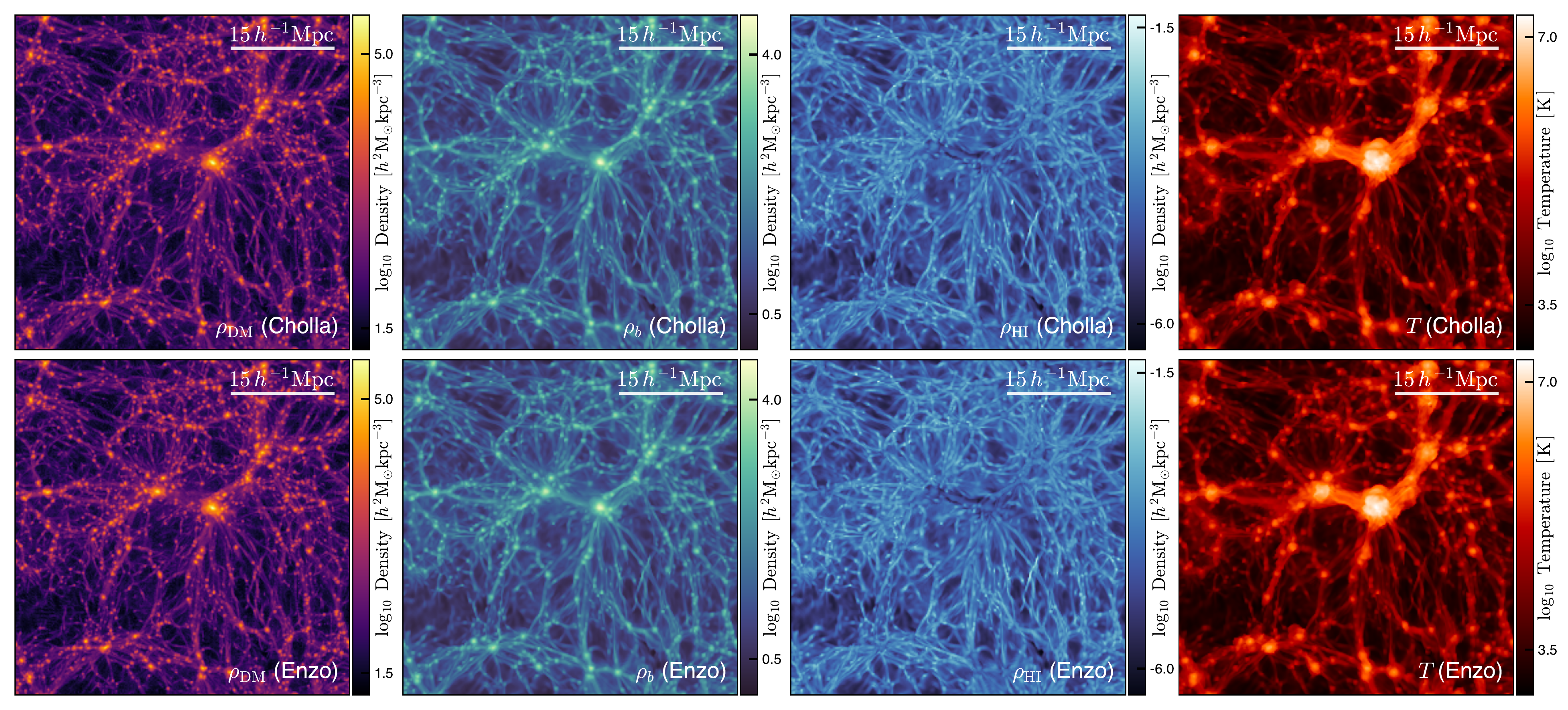}
\caption{Comparison of simulations (256$^3$ grid and $L=50h^{-1}\mathrm{Mpc}$) evolved with \Cholla (top panels) and \Enzo (bottom panels). From left to right the columns correspond to projections of dark matter density $\rho_{\mathrm{DM}}$, gas density $\rho_b$, neutral hydrogen density $\rho_{\mathrm{HI}}$, and gas temperature $T$, all at redshift $z=0$.}
\label{fig:full_hydro_projections}
\end{figure*}

\begin{figure*}[t!]
\includegraphics[width=\textwidth]{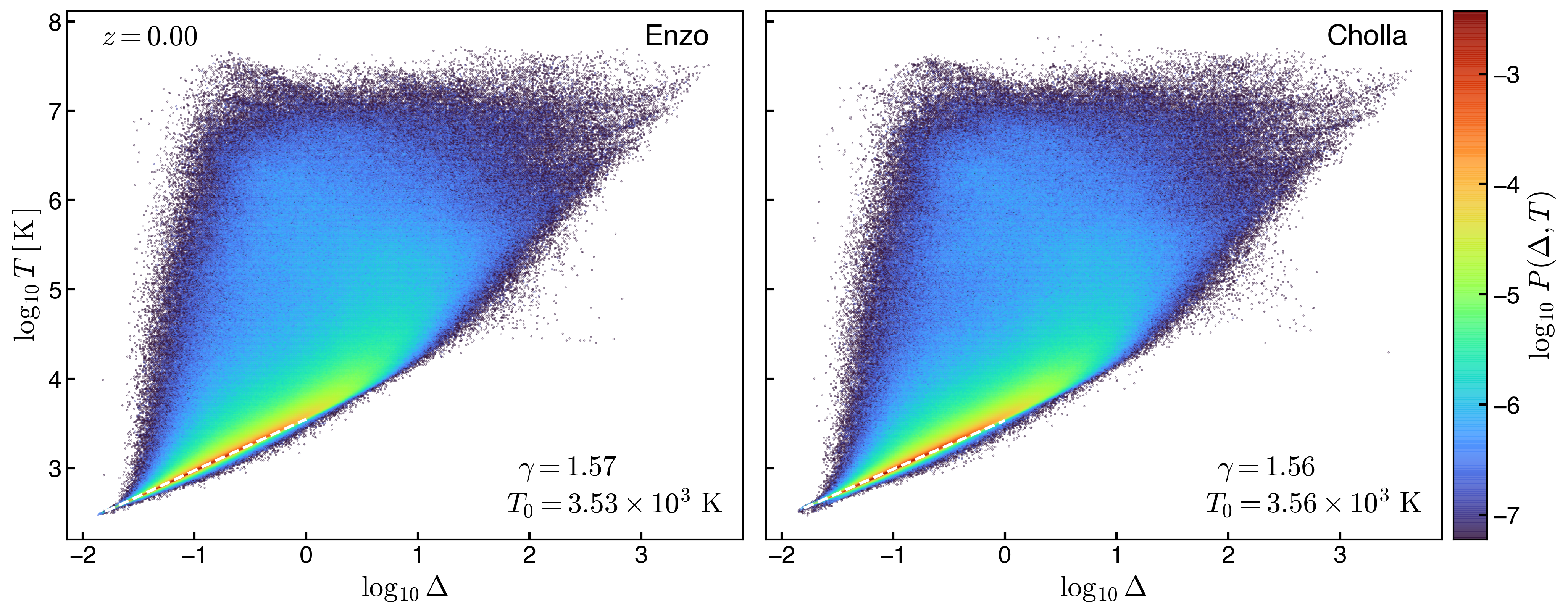}
\caption{Comparison of the density-temperature ($\Delta=\rho_b/ \bar{\rho_b}$) distribution  at redshift $z=0$ for analogous simulations (256$^3$ grid and $L=50h^{-1}\mathrm{Mpc}$) evolved with  \Enzo (left) and \Cholla (right) using the HM12 photoheating and photoionzation rates. The distribution of the gas in both simulations is remarkably similar and the differences for the parameters $T_0$ and $\gamma$ are $< 1\%$, demonstrating an excellent agreement for the gas distribution in the IGM between the codes. (The parameters $T_0$ and $
\gamma$ are defined in \S\ref{sec:igm_evolution}.)    }
\label{fig:phase_diagram_enzo_cholla}
\end{figure*}

The results of this comparison are shown in Figure \ref{fig:avrg_temperature},
where we plot the 
mass-weighted average temperature of the gas 
simulations evolved with \Ramses (purple line) or \Enzo (green line).
For \Cholla we ran two simulations with different dual energy conditions,
using a criteria similar to \Ramses (Equation \ref{eq:dual_energy_condition_ramses}, light blue line)
or similar to \Enzo (Equation \ref{eq:dual_energy_condition}, dark blue line).

All simulations start from the same initial temperature $T_0=230 \,\mathrm{K}$ at $z=100$.
Afterward, the gas cools as $T\propto a^{-2}$ owing
to universal expansion until the first halos collapse. The temperature of the infalling gas increases
owing to shock heating, causing the global average temperature to increase. 
As Figure \ref{fig:avrg_temperature} shows, the temperature increase happens at roughly two different times for the different simulations.
In the \Enzo calculation and corresponding \Cholla simulation that uses Equation \ref{eq:dual_energy_condition}
for the dual energy condition,
shock heating in the halos becomes significant at $z \sim 15$ (green and dark blue lines). 
In the \Ramses simulation and the \Cholla calculation using the condition given by Equation \ref{eq:dual_energy_condition_ramses} (purple and light blue lines), the gas continues cooling owing to expansion until $z \sim 6$. 

The delayed gas heating in calculations using the dual energy condition given by Eqn. \ref{eq:dual_energy_condition_ramses} results from the advected internal energy $e$ being dominantly selected over the
conserved internal energy $E-v^2/2$ for the gas infalling into halos.
In this case, the evolution of the advected internal energy $e$ is given by Equation \ref{eq:internal_energy_1}
that does not capture shock heating, and consequently the heating of the gas in the halos is suppressed.

In contrast, if Equation \ref{eq:dual_energy_condition} is used for the dual energy condition
in the \Cholla and \Enzo simulations,
the resulting mean cosmic temperature closely follows the virial temperature estimate at $z<6$.
We found that for \Chollan, adopting the parameter value $\eta=0.035$ in Equation \ref{eq:dual_energy_condition} results in a temperature increase that begins at $z\approx 12$, similar to our model estimate.

We note that the mean cosmic gas temperature measured 
in the Cholla adiabatic simulation and our $\bar{T}_{vir}$ estimate (Eq. \ref{eq:virial_temperature}) 
computed from the halo properties (Figure 5, dark blue and yellow lines respectively) display a sharp transition at $z \approx 12$, 
when the temperature suddenly increases. We argue that this behavior is a consequence of the limited resolution in our test simulation, 
as most of the low mass halos that would form at early times ($z>12$) are not resolved and the heating of gas during their
virialization is not captured.
We verify this by computing an estimate of $\bar{T}_{vir}$ including early, low-mass halos predicted 
by an analytical mass function \citep{Sheth&Tormen1999}. This
analytical estimate shows an earlier and more gradual increase in the mean temperature, starting at $z\sim 20$. If instead we 
limit the minimum halo mass used for the analytical estimate to the minimum resolved halo mass in the test simulations, 
the estimate follows closely the temperature evolution from the Cholla simulation including the sharp increase at $z\approx 12$,
strongly 
suggesting that the missing unresolved halos explain the sharp feature in the temperature evolution.

Additionally, we measured the effect on the gas overdensity power spectrum of temperature differences arising
from the choice of the dual energy condition.
Figure \ref{fig:ps_dual_energy_difference} shows the fractional difference of the power spectrum measured in the simulation where Equation \ref{eq:dual_energy_condition_ramses} was used for the dual energy selection relative to the power spectrum measured in the simulation that instead employed Equation \ref{eq:dual_energy_condition}.
The comparison shows that the power spectrum in the simulation using condition set by Eqn. \ref{eq:dual_energy_condition_ramses}, 
where the gas remains colder for longer, is $\sim 50\%$ higher on small scales by $z\sim5$. 
By $z\sim2$ differences on small scales reach $\sim 80\%$. Afterward,
the two simulations reach similar average temperatures and the differences decrease to $\sim 30\%$ by $z=0$.

We note that the truncation error $e_{\text{trunc}}$ (Eq. \ref{eq:truncation_error}) is resolution dependent, 
and the suppressed shock heating presented in this comparison might be
reduced in high resolution AMR simulations. Studying the behavior of condition \ref{eq:dual_energy_condition_ramses}
in such simulations is beyond the scope of this work.

\begin{figure*}[t!]
\includegraphics[width=\textwidth]{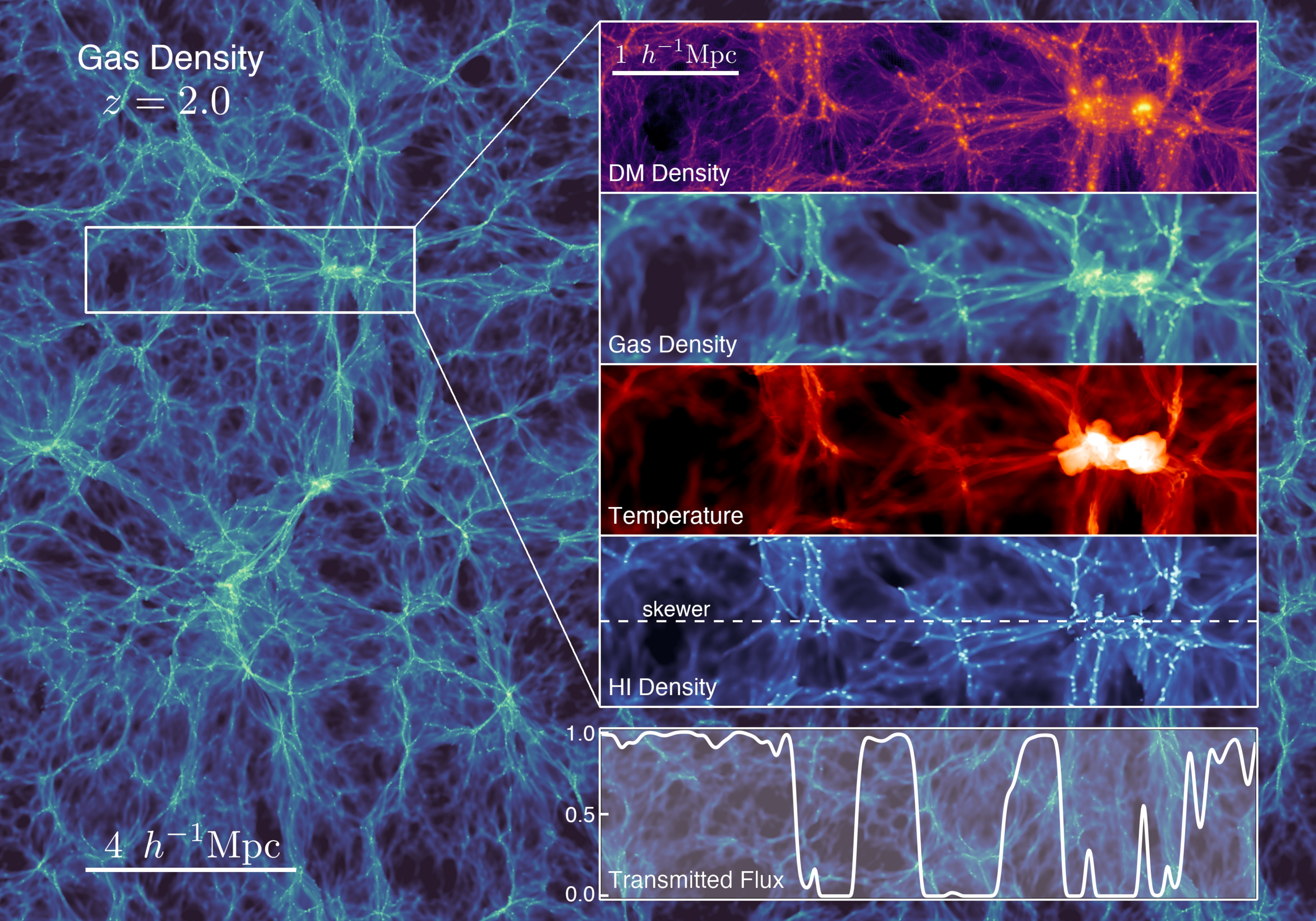}
\caption{Projection of the gas density at redshift $z=2$ from the \Simn.P19 simulation (2048$^3$ grid, $L=50 \,h^{-1}\mathrm{Mpc}$, and $1.5 \,h^{-1}\mathrm{Mpc}$ projected). The zoom-in region shows the dark matter density $\rho_{\mathrm{DM}}$, gas density $\rho_{b}$, gas temperature $T$, and neutral hydrogen density $\rho_{\mathrm{HI}}$ from top to bottom. A skewer crossing the center of the zoom-in region is marked over the neutral hydrogen distribution, and the Lyman-$\alpha$ transmitted flux along the skewer is shown in the bottom panel. The scale labels refer to proper distances.}
\label{fig:high_res_hydro_projection}
\end{figure*}

\subsection{Cosmological Simulation: Chemistry and UV Background}
\label{sec:chemistry_test}

To validate our integration of the chemical network (solved by GRACKLE and advected by the hydrodynamics solver), we ran identical hydrodynamical simulations solved with \Enzo and \Cholla following the configuration described in
\S \ref{sec:adiabatic_hydro_sim} but including the non-equilibrium H and He network plus metals in the presence of a spatially-uniform, time-dependent UV background given by the standard HM12 \citep{haardt2012a} photoheating and photoionization rates.
Figure \ref{fig:full_hydro_projections} shows a comparison of projected gas quantities at $z=0$ computed by the
two codes, with
\Cholla on top and \Enzo on the bottom. From left to right the panels show projections of dark matter density ($\rho_{\mathrm{DM}}$), gas density ($\rho_b$), neutral hydrogen density ($\rho_{\mathrm{HI}}$), and gas temperature ($T$). Figure \ref{fig:full_hydro_projections} shows qualitatively a remarkable agreement between the results from the two codes. 

For a more detailed comparison, we measured the density-temperature distribution from both simulations. Figure \ref{fig:phase_diagram_enzo_cholla} shows the results from the \Enzo (left) and \Cholla (right) runs at $z=0$. As shown, the distribution of the gas in the simulations is remarkably similar. Additionally the parameters $T_0$ and $\gamma$ which model the distribution of the diffuse gas in the IGM (see \S\ref{sec:igm_evolution} for details) differ by less than 1\% showing an excellent agreement between these simulations.

\section{Simulation Suite}
\label{sec:simulation_suite}

In this section we present the \Sim (\textbf{CH}olla \textbf{I}GM \textbf{P}hothoheating \textbf{S}imulations) simulation suite, a set of high resolution simulations performed using the newly extended version of the \Cholla code described above. The suite consists of a series
of simulations run with a fiducial resolution of $N=2048^3$ cells, varying the cosmic photoheating and photoionization rates from evolving UV background radiation fields and with a range of cosmological parameters. All the \Sim simulations evolve a primordial gas composition ($X=0.76$, $Y=0.24$), without the inclusion of metal line cooling as star formation is not accounted for in the simulations. 

\begin{deluxetable*}{lcccc}
\tablenum{1}
\caption{\Sim Simulation Suite\label{tab:sims}}
\tablewidth{\linewidth}
\tablehead{
	\colhead{Simulation} & \colhead{Resolution} & \colhead{Box Size} & \colhead{Parameters} & \colhead{UV Background}\\[-6pt]
	\colhead{} & \colhead{} & \colhead{$L~[h^{-1}\mathrm{Mpc}]$} & \colhead{[$h$, $\Omega_{m}$, $\Omega_{b}$, $\sigma_8$, $n_s$]} & \colhead{}
}
\startdata
\Simn.HM12 & $N=2048^3$ & $50$ & $[0.6766,0.3111,0.0497,0.8102,0.9665]$ & \citet{haardt2012a}\\
\Simn.P19 & $N=2048^3$ & $50$ & $[0.6766,0.3111,0.0497,0.8102,0.9665]$ & \citet{puchwein2019a}\\
\hline
\multicolumn{5}{c}{Alternative Cosmologies}\\
\hline
\Simn.P19.A1 & $N=2048^3$ & $50$ & $[0.6835,0.3010,0.0484,0.8098,0.9722]$ & \citet{puchwein2019a}\\
\Simn.P19.A2 & $N=2048^3$ & $50$ & $[0.6917,0.2905,0.0477,0.8052,0.9783]$ & \citet{puchwein2019a}\\
\Simn.P19.A3 & $N=2048^3$ & $50$ & $[0.7001,0.2808,0.0470,0.8020,0.9846]$ & \citet{puchwein2019a}\\
\Simn.P19.A4 & $N=2048^3$ & $50$ & $[0.7069,0.2730,0.0465,0.7997,0.9896]$ & \citet{puchwein2019a}\\
\hline
\multicolumn{5}{c}{Resolution Studies}\\
\hline
\Simn.P19.R1 & $N=1024^3$ & $50$ & $[0.6766,0.3111,0.0497,0.8102,0.9665]$ & \citet{puchwein2019a}\\
\Simn.P19.R2 & $N=512^3 $ & $50$ & $[0.6766,0.3111,0.0497,0.8102,0.9665]$ & \citet{puchwein2019a}\\
\enddata
\tablecomments{Resolution refers to the number of both grid cells and dark matter particles.}
\end{deluxetable*}

Table \ref{tab:sims} details the properties our initial \Sim simulations. 
The primary simulations for our initial
analysis of the \Lya forest are \Simn.HM12 and \Simn.P19, which use the \citet{planck2018a}
cosmological parameters and the \citet{haardt2012a} and \citet{puchwein2019a} photoionization and
photoheating rates, respectively, in an $L=50~h^{-1}\mathrm{Mpc}$ box. 
The \citet{puchwein2019a} model adopts the most recent determinations of the ionizing emissivity due to stars and AGN, as well as of the \HI absorber column density distribution. Another major improvement is a new treatment of the IGM opacity for ionizing radiation that is able to consistently capture the transition from a neutral to ionized IGM.   
For these fiducial runs, we 
output 150 snapshots over the redshift range $z=[16,2]$, spacing the time between
snapshots at $\Delta a = 1.83 \times 10^{-3}$ intervals. In each snapshot, we record the
conserved fluid quantities ($\rho$, $\rho v_{x}$, $\rho v_{y}$, $\rho v_{z}$, $E$),
the gas internal energy $u$,
the neutral hydrogen \ion{H}{1}, neutral helium \ion{He}{1}, 
singly-ionized helium \ion{He}{2}, and electron $n_{e}$ densities, 
and the gravitational potential $\phi$ on the simulation grid. We also record all the
dark matter particle positions and velocities.
The detailed analyses performed on the
simulation outputs are described in \S \ref{sec:igm_evolution}.

We complement the fiducial models
with four additional simulations (\Simn.P19.[A1-A4]) that use the \citet{puchwein2019a}
photoionization and photoheating rates but vary the cosmological parameters $h$, $\Omega_m$, $\Omega_b$,
$\sigma_8$, and $n_s$ within the uncertainties reported by \citet{planck2018a}. For each simulation,
a flat cosmology is assumed and we set $\Omega_{\Lambda} = 1-\Omega_m$.

Table \ref{tab:sims} also lists properties of
the additional $N=1024^{3}$ (\Simn.P19.R1) and $N=512^{3}$ (\Simn.P19.R2)
simulations used in our resolution study to demonstrate the numerical convergence of our results
(see Appendix \ref{sec:resolution}).

The \Sim simulation suite was run on the Summit system (Oak Ridge Leadership Computing Facility at the Oak Ridge National Laboratory), each of the $2048^3$ simulations ran in 512 GPUs for $\sim$11 hours costing $\sim$1000 node-hours. As described in \S \ref{sec:algorithm}, the slowest component of the simulation is the GRACKLE update step, consuming about half of the computational time for these runs. This motivates the ongoing development of a H+He network solver implemented to run in the GPUs which will potentially reduce this bottleneck. The subsequent analysis of the simulation output data was performed using the \emph{lux} supercomputer at UC Santa Cruz.

\section{Evolution of the IGM for Two Photoheating Histories}
\label{sec:igm_evolution}

Redshift-dependent photoionization and photoheating rates of intergalactic
gas substantially affect IGM properties. By comparing the \Simn.HM12 with the \Simn.P19 simulation, we can learn about how 
detailed differences in photoheating history lead to observable
differences in the \Lya forest and potentially discriminate between
them by further comparisons with data. We first
compare the thermal history of the diffuse IGM between the
simulations (\S \ref{sec:thermal_history}). The redshift-dependent
thermal properties of the IGM in the models provide a context for
interpreting measurements of the simulated forest. We discuss our
methods for generating mock \Lya absorption spectra in \S
\ref{sec:synthetic_spectra_skewer}. These simulated spectra then
provide estimates of the \Lya forest optical depth
(\S \ref{sec:optical_depth}) and transmitted flux power spectra
(\S \ref{sec:power_spectra}).

\begin{figure*}[t!]
\includegraphics[width=\textwidth]{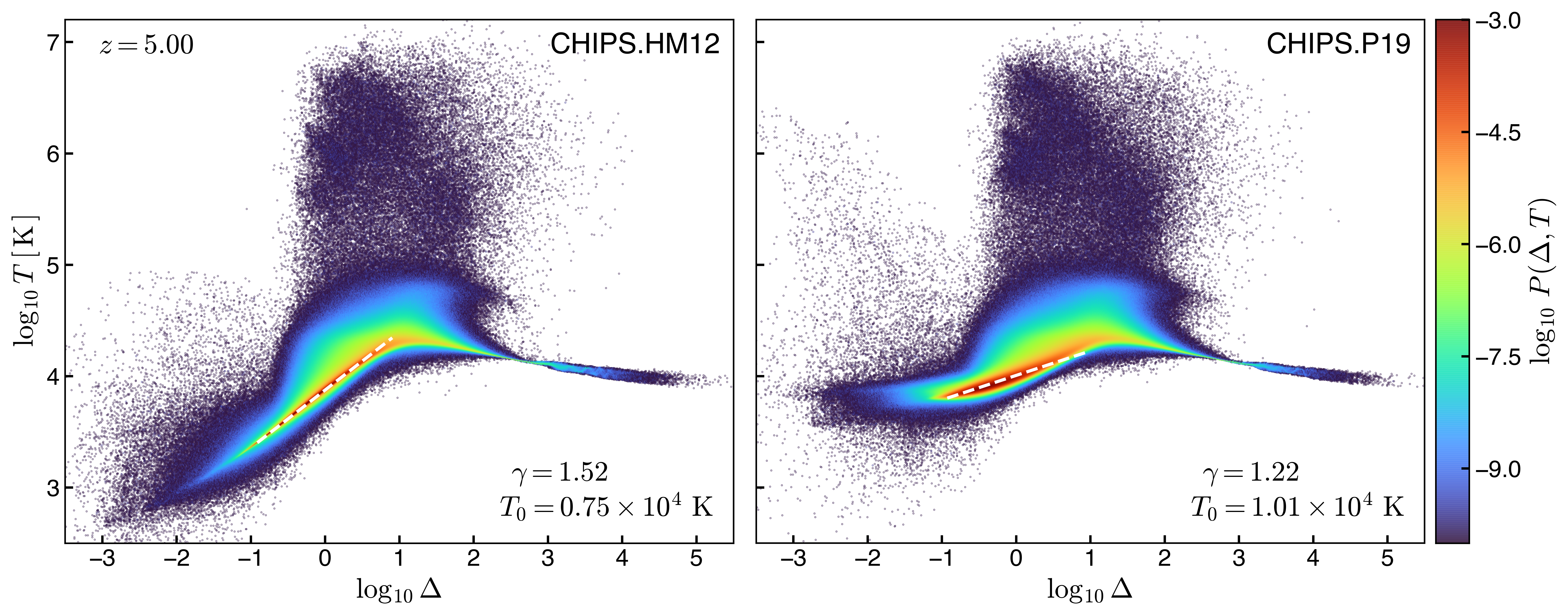}
\caption{Volume-weighted density-temperature distribution of gas at redshift $z=5$ in our two fiducial simulations (\Simn.HM12, left; \Simn.P19, right). The low density gas ($\Delta < 10$) is colder for the \cite{haardt2012a} model owing to $\mathrm{H}$ reionization ending earlier. Dashed lines show the best fit power-law $\Delta$-$T$ relation for the parameters $T_0$ and $\gamma$.}
\label{fig:phase_diagram_2048}
\end{figure*}

\subsection{Thermal History of the Diffuse IGM}
\label{sec:thermal_history}

The gas in the diffuse IGM comprises most of the baryons in the universe and
follows a well defined density-temperature power-law relation
\citep{hui1997a, mcquinn2016a, puchwein2015a} given by

\begin{equation}
T(\Delta) = T_0 \Delta ^{\gamma - 1},
\label{eq:rho_T_relation}
\end{equation} 
\noindent
where $\Delta=\rho_b/ \bar{\rho_b}$ is the gas overdensity, $T_0$ is the temperature at the
mean cosmic density $\bar{\rho_b}$, and $\gamma-1$ corresponds to the power-law index of the relation. The time evolution of the parameters $T_0$ and $\gamma$ is determined by the photoheating from to hydrogen and helium ionization,
cooling owing to the expansion of the universe, and
inverse Compton cooling, recombination, and collisional processes. 

Figure \ref{fig:phase_diagram_2048} shows the density-temperature distribution of the gas in our simulations at
redshift $z=5$, with \Simn.HM12 shown on the left and \Simn.P19 shown on the right.
The distributions resulting from the two UVB models are similar for gas collapsed into resolved
structures $(\Delta>10)$, but for low density gas $(\Delta<10)$ the temperatures in the HM12 model are significantly lower owing to the earlier completion of hydrogen reionization $(z\sim13)$.
The gas temperature in this model has had subsequently more time to decrease owing to
cooling processes and adiabatic expansion.
For the P19 model, where reionization ends at $z\sim 6$, there has been less time to cool by $z=5$,
resulting in a higher $T_0$ and lower $\gamma$ at this epoch.      

\begin{figure*}
\includegraphics[width=\textwidth]{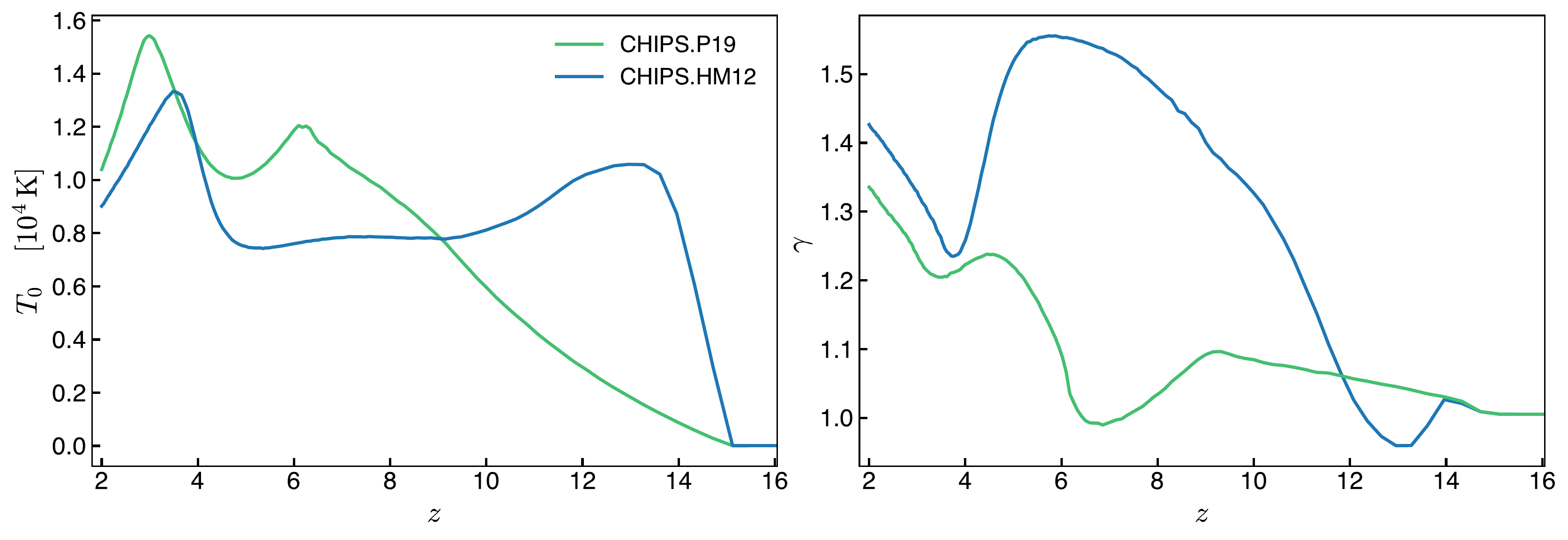}
\caption{Redshift evolution of the gas temperature $T_0$ at mean density and the index $\gamma$ of the density-temperature relation.
Shown are results from our reference simulations \Simn.HM12 (blue lines) and \Simn.P19 (green lines). \HI reionization ends earlier ($z\sim 13$) for the HM12 model, compared with $z\sim 6$ for the P19 model, allowing more time for the low density gas to cool. \HeII reionization begins at $z \sim 4.5$ in both models, but the lower \HeII photoionization rates for the P19 model result in \HeII being fully ionized at a later time ($z\sim 3$) compared with $z\sim 3.8$ for the HM12 run.      }
\label{fig:thermal_history}
\end{figure*}

For each snapshot of the simulation, we determined the parameters $T_0$ and $\gamma$
by fitting Equation \ref{eq:rho_T_relation} 
to the low density ($-1<\log_{10}\Delta<1$) region of the density-temperature distribution. 
We divided the selected interval into fifty equal bins in $\log_{10} (\Delta)$, and for
each bin $i$ the maximum of the marginal temperature distribution $P(T | \Delta_i)$ and
the temperature range containing the 68\% highest probability density 
was used to define the bin temperature $T_i$ and its corresponding uncertainty $\delta T_i$.
The coordinates ($\Delta_i$, $T_i$) and uncertainty values $\delta T_i$ were used as input to a 
Monte Carlo Markov Chain that sampled the parameters $T_0$ and $\gamma$, initialized from uniform prior distributions, and returned posterior distributions for the thermal parameters that best match the density-temperature relation measured from the simulations.
From the posterior distributions we extracted best-fit values for $T_0$ and $\gamma$ and corresponding
parameter uncertainties $\delta T_0$ and $\delta \gamma$.

\begin{figure*}
\includegraphics[width=\textwidth]{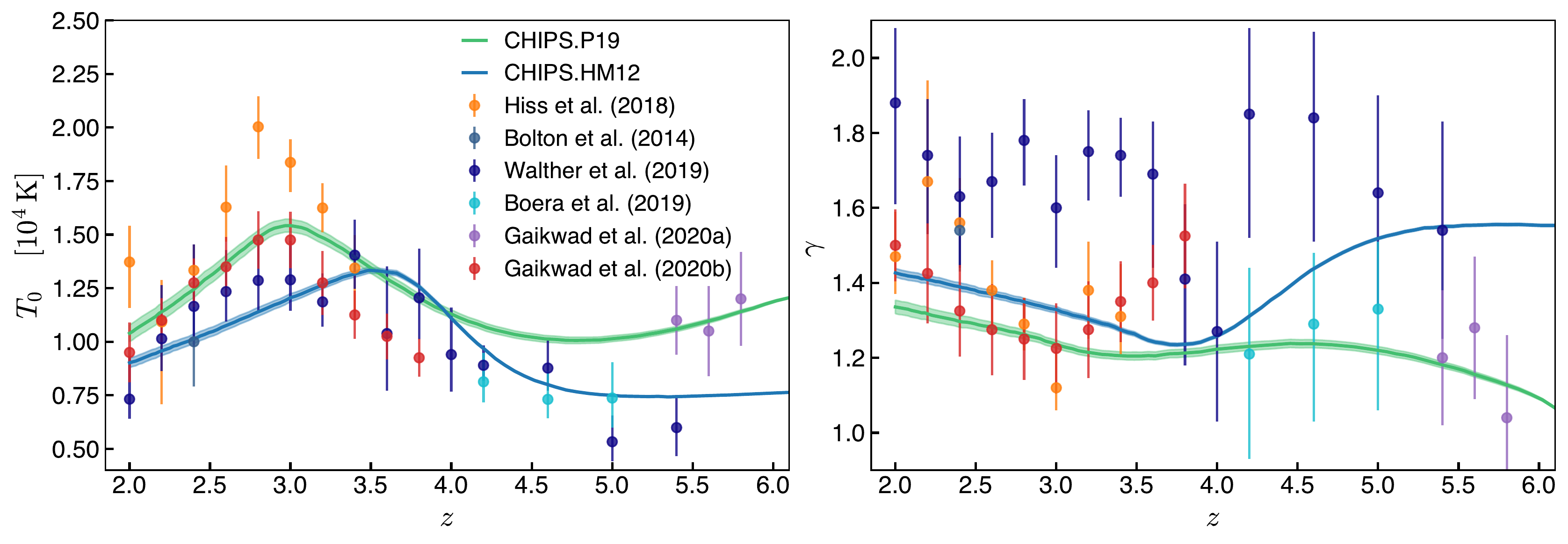}
\caption{Redshift evolution of the density-temperature parameters $T_0$ and $\gamma$ from the \Simn.HM12 (blue) and \Simn.P19 (green)
simulations. The shaded regions show the uncertainty resulting from the power-law fitting procedure. Points show the 
observational
results from \cite{hiss2018a}, \cite{bolton2014a}, \cite{walther2019a}, \cite{boera2019a}, \cite{gaikwad2020a}, and \cite{gaikwad2020b}.}
\label{fig:thermal_history_data}
\end{figure*}

The redshift evolution of the parameters $T_0$ and $\gamma$ for the two UVB models is shown in Figure \ref{fig:thermal_history}, where the effects of the two ionization histories on the thermal structure of the diffuse IGM are illustrated. The evolution of the IGM follows several phases. First, before \HI reionization
is complete, the photoheating owing to hydrogen ionization increases the temperature to an early local maximum.
After all the \HI is ionized, the diffuse IGM cools by adiabatic expansion and inverse Compton
cooling until the onset of \HeII reionization reheats the IGM to a global maximum. Once \HeII is fully
ionized, the IGM again cools adiabatically to the present day.

The HM12 UVB (blue line) causes a quick \HI reionization around redshift $z\sim14$ and cools afterward.
In the interval between $z\sim6-10$, the temperature at mean density $T_0$ plateaus at $T\sim 8 \times 10^3 \, \mathrm{K}$ while underdensities $(\Delta<1)$ keep cooling, mostly due to adiabatic expansion, this
results in an increasing $\gamma$ until $z \sim 5$. 

For $z\leq 15 $ the ionization rates for the late-reionization P19 model are significantly lower until $z\sim 6.5$, 
resulting in a gradual heating of the IGM. The IGM remains close to isothermal ($\gamma \sim 1$) until
\HI reionization completes at $z\sim 6.2$. Intergalactic gas then cools just for short period before \HeII reheating, resulting in a higher $T_0$ than in the HM12 model. 

While in both runs helium reheating starts around $z \sim 4.5$, the reionization of \HeII is completed earlier in \Simn.HM12. At $ z< 3$, the residual heating from photoionization of recombining atoms is inefficient, and the IGM continues to cool all the way down to $z=0$, decreasing $T_0$ and increasing $\gamma$.

Figure \ref{fig:thermal_history_data} shows a comparison between the thermal parameters $T_0(z)$ and $\gamma(z)$ from our simulations and previous observational inferences \citep[]{bolton2014a,hiss2018a,boera2019a,walther2019a,gaikwad2020a,gaikwad2020b}.
The shaded regions correspond to the uncertainty in $T_0$ and $\gamma$ resulting from our power-law fitting procedure.
For the observations, the values of $T_0$ and $\gamma$ are determined in different ways.

\cite{walther2019a} and \cite{boera2019a} both follow a similar approach by generating Lyman-$\alpha$ flux power spectra from simulations evolved with different thermal histories,
resulting in multiple trajectories of $T_0$ and $\gamma$.
For each simulation snapshot, they determine the best fit $T_0$, $\gamma$, and mean transmitted flux $\bar{F}$
by performing Bayesian inference comparing the generated flux power spectra from the different simulations to observations. 

\cite{hiss2018a} and \cite{bolton2014a} measure the $b$-$N_{\mathrm{HI}}$ distribution obtained from decomposing the Lyman-$\alpha$ forest spectra into a collection of Voigt profiles, and then infer thermal parameters 
by matching the $b$-$N_{\mathrm{HI}}$ distribution from their simulations to the observed distribution. 
\cite{gaikwad2020b} follows an analogous approach by comparing simulation results to Voigt profiles fitted
to transmission spikes in the inverse transmitted flux $1-F$ in $z>5$ spectra.
In a recent analysis, \cite{gaikwad2020a} report more precise results by inferring $T_0$ and $\gamma$ from combined constraints  
obtained through a comparison between simulated and observed \Lya forest flux power spectra,
$b$-$N_{\mathrm{HI}}$ distributions, wavelet statistics, and curvature statistics.

During the epoch of \HeII reionization and afterward, 
\cite{hiss2018a} infer a peak in $T_0$ ($z \sim 2.8$) that is significantly higher than the results from all the other analysis, while the measurements from \cite{gaikwad2020a} are mostly higher than those obtained by \cite{walther2019a} and \cite{bolton2014a} their results still  are consistent within $1\sigma$ of each other. Compared to the simulations, both the P19 and HM12 models result in a increase of $T_0$ due to \HeII photoheating that begins too early ($z\sim 4.2$) to be consistent with the measurements from \cite{gaikwad2020a} and \cite{walther2019a} simultaneously. The heating from \HeII reionization could be delayed in the P19 model by decreasing the \HeII photoheating and photoionization rates, effectively also slightly decreasing the peak of $T_0$ at $z\sim 2.8$, this would produce a trajectory for $T_0$ at $z \lesssim 4$ that better matches the results from \cite{gaikwad2020a} and \cite{walther2019a}.         

For $z>4$, the results from \cite{walther2019a} and \cite{boera2019a} measure temperatures lower than those produced by the \cite{puchwein2019a} model. In particular,
at $z\geq5$ \cite{walther2019a} find remarkably low temperatures, which may be related to the strong correlation between $T_0$ and $\bar{F}$ in their analysis.

The $\gamma$ values inferred by \cite{walther2019a} are overall higher than all the other data sets.
In the redshift range  $ 2.6 \lesssim z \leq 5 $,
both our \Simn.HM12 and \Simn.P19  simulations produce $\gamma(z)$ consistent with the measurements 
from \cite{hiss2018a} and \cite{boera2019a}, within their respective uncertainties. Additionally both models result in $\gamma(z)$ consistent with the results from \cite{gaikwad2020a} only for $z \lesssim 3.3$ as their observational inference results in higher values ($\gamma \sim 1.5$) at $z \sim 3.8$  compared to $\gamma \sim 1.3$ produced in both simulations at this redshift. Delaying the heating from \HeII reionization would allow more time for the diffuse gas to cool after \HI reionization, effectively increasing $\gamma$ produced by the models at $z\sim 4$, to be in better agreement with the results from \cite{gaikwad2020a}.

The \Simn.P19 results are consistent with constraints on $T_0$ and $\gamma$ for $z > 5.3$ from \cite{gaikwad2020b},
likely because \HI reionization ends at $z\sim 6$ in the \cite{puchwein2019a} photoionization and photoheating model. Although, reconciling the high $\gamma \sim 1.5 $ at $z \sim 3.8$ from \cite{gaikwad2020a}  with the low $\gamma \sim 1.2$ at $z \sim 5.5$ inferred by \cite{gaikwad2020b}  could require the low density gas to cool faster than physically possible in a spatially-uniform UV background model.
However, near \HI reionization a non-uniform UV background may be required for the simulations to model 
accurately the effects of a ``patchy'' reionization \citep{keating2020a}.

\subsection{Synthetic \Lya Forest Spectra }
\label{sec:synthetic_spectra_skewer}

\begin{figure*}[t!]
\includegraphics[width=\textwidth]{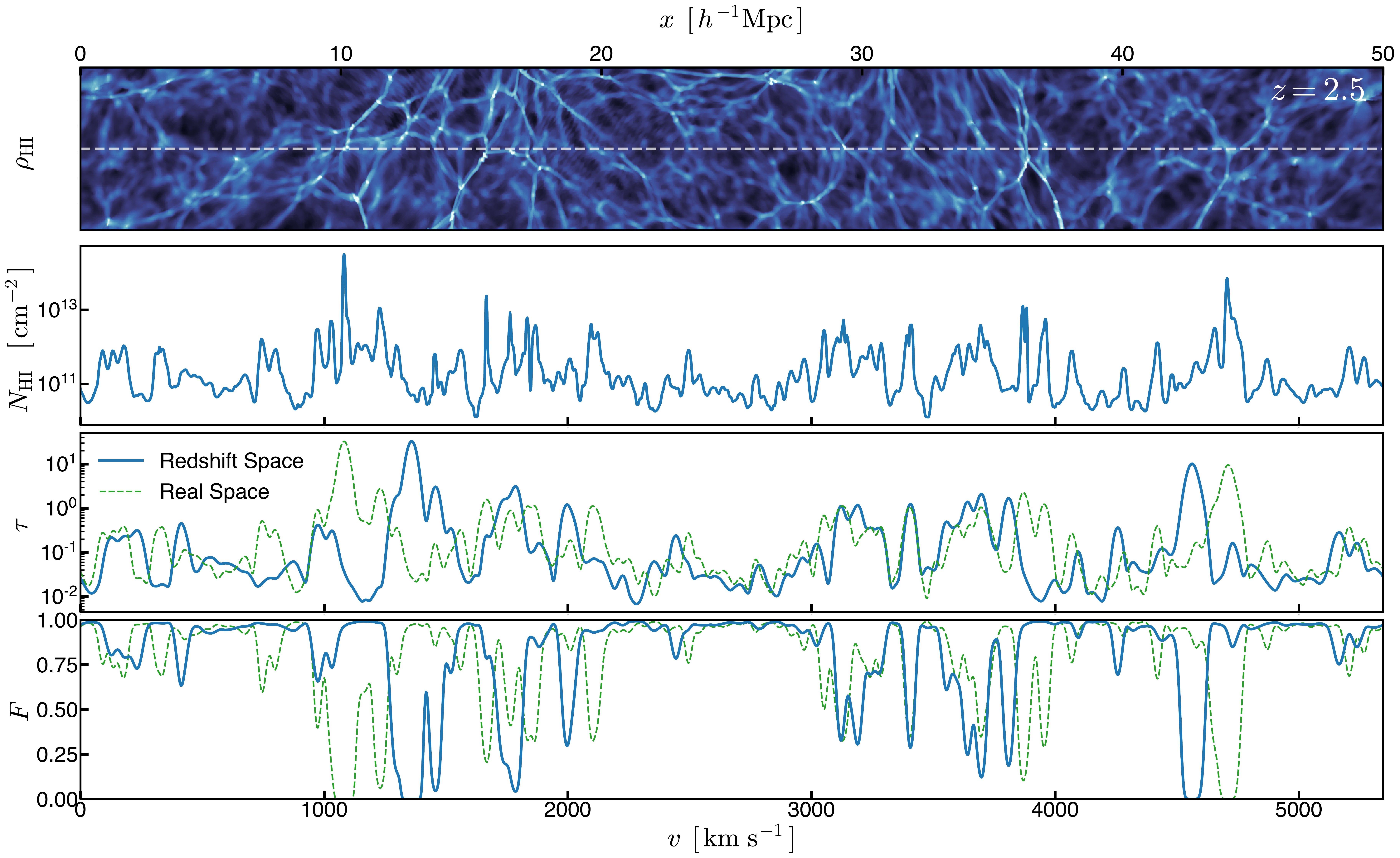}
\caption{Elements for the calculation of the transmitted flux along a single skewer crossing the \Simn.P19 simulation box at redshift $z=2.5$. Shown in the panels from top to bottom are the neutral hydrogen density $\rho_{\mathrm{HI}}$ in a region surrounding the skewer, neutral hydrogen column density $N_{\mathrm{HI}}$ along the skewer, optical depth $\tau$ computed via Equation \ref{eq:optical_depth_error_function}, and transmitted flux $F=\exp (-\tau)$ along the line of sight.}
\label{fig:transmitted_flux_skewer}
\end{figure*}

The Lyman-$\alpha$ forest is a sensitive probe of the diffuse baryons in the IGM, 
as the amplitude and width of the absorption lines in the forest trace the neutral hydrogen density and the gas temperature.
The observed
Lyman-$\alpha$ forest global statistics, such as the effective optical depth and the transmitted flux power spectra,
constrain the thermal state of the IGM. 
To compare directly with the observed effective optical depth and the transmitted flux power spectra,
we compute synthetic Lyman-$\alpha$ forest spectra from our high resolution simulations.
In total we drew 60000 skewers through the simulation volume, located in random positions and aligned parallel
with the three box axes (20000 skewers for each axis). 
Along each skewer, the neutral hydrogen density, gas temperature, and the component of the velocity parallel to the line of sight are sampled at the native resolution of the simulation,
rendering 2048 uniformly distributed pixels for each skewer.
The optical depth as a function of frequency $\tau_{\nu}$ along the skewer is computed by integrating the Lyman-$\alpha$ interaction cross section $\sigma_{\nu}$ and the neutral hydrogen number density $n_{\mathrm{HI}}$ along the line of sight, following
\begin{equation}
\tau_{\nu}=\int n_{\mathrm{HI}} \sigma_{\nu} d r,
\end{equation}
\noindent
where $dr$ is the physical length of the path element.
Assuming a Doppler profile for the absorption line, the optical depth at frequency $\nu_0$ is given by
\begin{equation}
\tau_{\nu_0}=\frac{\pi e^{2}}{m_{e} c} f_{12} \int \frac{n_{\mathrm{HI}}}{\sqrt{\pi} \Delta \nu_{\mathrm{D}}} \exp \left[-\left(\frac{\nu-\nu_0}{\Delta \nu_{\mathrm{D}}}\right)^{2}\right] d r, 
\end{equation}
\noindent
where $f_{12}$ is the Lyman-$\alpha$ transition upward oscillator strength, $\Delta \nu_{\mathrm{D}} = (b/c) \nu_0$ is the absorption width owing to Doppler shifts and $b=\sqrt{2 k_{\mathrm{B}} T / m_{\mathrm{H}}}$ corresponds to the thermal velocity of the gas. The shift in the frequency of absorption along the skewer is given by the Doppler shift from the change in the gas velocity along the line of sight,
\begin{equation}
\nu=\nu_{0}\left(1-\frac{u - u_0}{c}\right).
\end{equation} 
\noindent
Applying a variable transformation from frequency to velocity space and the expansion relation $d u = H d r$, the optical depth as a function of velocity is expressed as
\begin{equation}
\tau_{u_0}=\frac{\pi e^{2} \lambda_0}{m_{e} c H} f_{12} \int \frac{n_{\mathrm{HI}}}{\sqrt{\pi} b} \exp \left[-\left(\frac{u-u_0}{b}\right)^{2}\right] d u.
\end{equation}
\noindent
Following the method described by \cite{lukic2015a}, we solved the
Gaussian integral analytically and computed the optical depth along the discretized line of sight using
\begin{equation}
\tau_{j}=\frac{\pi e^{2} \lambda_{0} f_{12}}{m_{e} c H} \sum_{i} \frac{1}{2}n_{\mathrm{HI}, i} \left[ \operatorname{erf}\left(y_{j+1 / 2, i}\right)-\operatorname{erf}\left(y_{j-1 / 2, i}\right) \right],
\label{eq:optical_depth_error_function}
\end{equation}  
\noindent
where the argument to the error function is
\begin{equation}
y_{j \pm 1 / 2, i}=\left(v_{\mathrm{H},j \pm 1 / 2}-v_{\mathrm{H}, i}-v_{\mathrm{LOS}, i}\right) / b_{i}.
\end{equation}
\noindent
The term $v_{\mathrm{H},j \pm 1 / 2}$ corresponds to the Hubble flow velocity at the interfaces of cell $j$ and the terms  $v_{\mathrm{H}, i}$ and  $v_{\mathrm{LOS}, i}$ represent the centered values of Hubble velocity and the line of sight component of the peculiar velocity of the gas at cell $i$.
Note that the factor of $1/2$ in Equation \ref{eq:optical_depth_error_function} comes from the definition used for the error function, $ \operatorname{erf}(x) = 1/\sqrt{\pi}\int_{-x}^{x}\exp (-t^2) dt$.

The calculation of the optical depth $\tau$ and the transmitted flux for a single skewer spanning
across the length of the \Simn.P19 box at redshift $z=2.5$ is illustrated in Figure \ref{fig:transmitted_flux_skewer}.
The figure panels from top to bottom show the distribution of the neutral hydrogen density in the neighborhood of the skewer ($\rho_{\mathrm{HI}}$), the 1D neutral hydrogen column density integrated over the length of the cell across the skewer ($N_{\mathrm{HI}}$), the optical depth $\tau$ in redshift space computed via Equation \ref{eq:optical_depth_error_function} (blue line) and ignoring the shift of of the absorption lines due to peculiar real space velocities (green), and the transmitted flux $F=\exp(-\tau)$ along the skewer in both redshift (blue) and real (green) space.

\subsection{Evolution of the \Lya Effective Optical Depth}
\label{sec:optical_depth}

\begin{figure}
\includegraphics[width=0.47\textwidth]{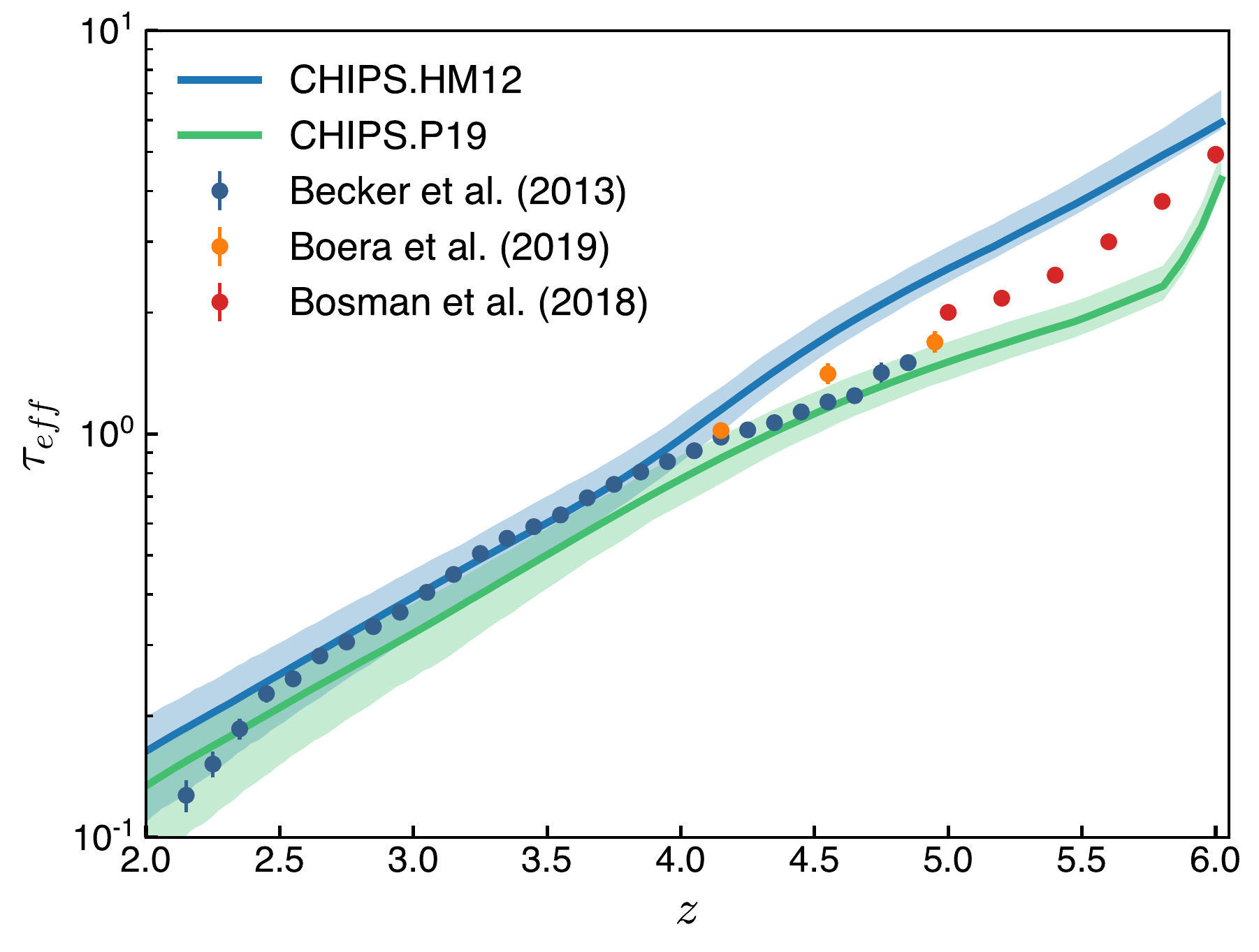}
\caption{Redshift evolution of the effective optical depth $\tau_{eff}$. Shown are our simulated measurements from \Simn.HM12 (blue) and  \Simn.P19 (green), compared to data from \cite{Becker_2013}, \cite{boera2019a} and \cite{Bosman_2018} (data points).
Measurements of $\tau_{eff}$ from the HM12  model match the observations at $2.5 \lesssim z \lesssim 4.2$, but display higher amounts of \HI for $ z \gtrsim 4.2$ owing to the low temperatures at those redshifts.
The P19 model produces values of $\tau_{eff}$ slightly lower than the observations for $2.5 \lesssim z \lesssim 4$, suggesting that \HeII reionization is overheating the IGM at these epochs. For $ 5 < z < 5.8$ the P19 model results in measurements of $\tau_{eff}$ significantly lower than the observations. This discrepancy could result from the
relatively hot IGM produced by the P19 model at this epoch and may be addressed by introducing a non-uniform UVB in the simulations.              }
\label{fig:optical_depth_redshift}
\end{figure}

The \Lya effective optical depth $\tau_{eff}$ is a measure of the overall \HI content of the gas in the IGM.
Hence, $\tau_{eff}$ tracks the ionization state of hydrogen and the intensity of the ionizing UV background.
To compare with observational measurements of $\tau_{eff}$, we computed synthetic \Lya absorption spectra from all the outputs of our two simulations using the method described in \S \ref{sec:synthetic_spectra_skewer}. From the large sample of skewers,
the effective optical depth is computed as $\tau_{eff}= - \log(\bar{F})$, where $\bar{F}$ is the transmitted flux averaged over all skewers. 

The redshift evolution of the effective optical depth $\tau_{eff}$ for our two simulations is shown as colored lines in Figure \ref{fig:optical_depth_redshift} (blue and green for the HM12 and P19 models, respectively) and the shaded region shows the variability of $\tau_{eff}$ measured over the different skewers.
For each redshift the shaded interval corresponds to the optical depth computed from the highest probability interval that encloses 68\% of the distribution of the transmitted flux averaged over individual skewers.

The effective optical depth resulting from the HM12 UVB model (blue) shows good agreement with the observed data
\citep{Becker_2013} for $z<4$, but underestimates the ionization fraction at $z>4$. At the higher redshifts, the
model produces an excess of neutral hydrogen, likely because the early \HI reionization renders the gas too cold, this results in an effective optical depth higher than estimated by \cite{boera2019a} and measured by \cite{Bosman_2018}. 

The P19 model (green) has too high an ionization fraction,
resulting in an optical depth that is slightly lower than the
observations.
In the redshift range $2.5 \lesssim z \lesssim 4$,
\HeII reionization in the P19 model is overheating the IGM.
For $ 5 < z < 5.8$, the P19 model produces $\tau_{eff}$ that are 10 to 25\% lower than the observations,
suggesting that the temperatures of the IGM in this model at $z\sim 5.2$ are higher than those in reality.
At these high redshifts ($z\gtrsim 5.4$), a non-uniform UV background may be
required to accurately represent the effects of a ``patchy'' reionization in  $\tau_{eff}$ \citep{keating2020a},
and the inclusion of a non-uniform UVB could reduce the discrepancies between the data and the P19 model
at these times.

\subsection{\Lya Transmitted Flux Power Spectrum}
\label{sec:power_spectra}

\begin{figure*}
\includegraphics[width=\textwidth]{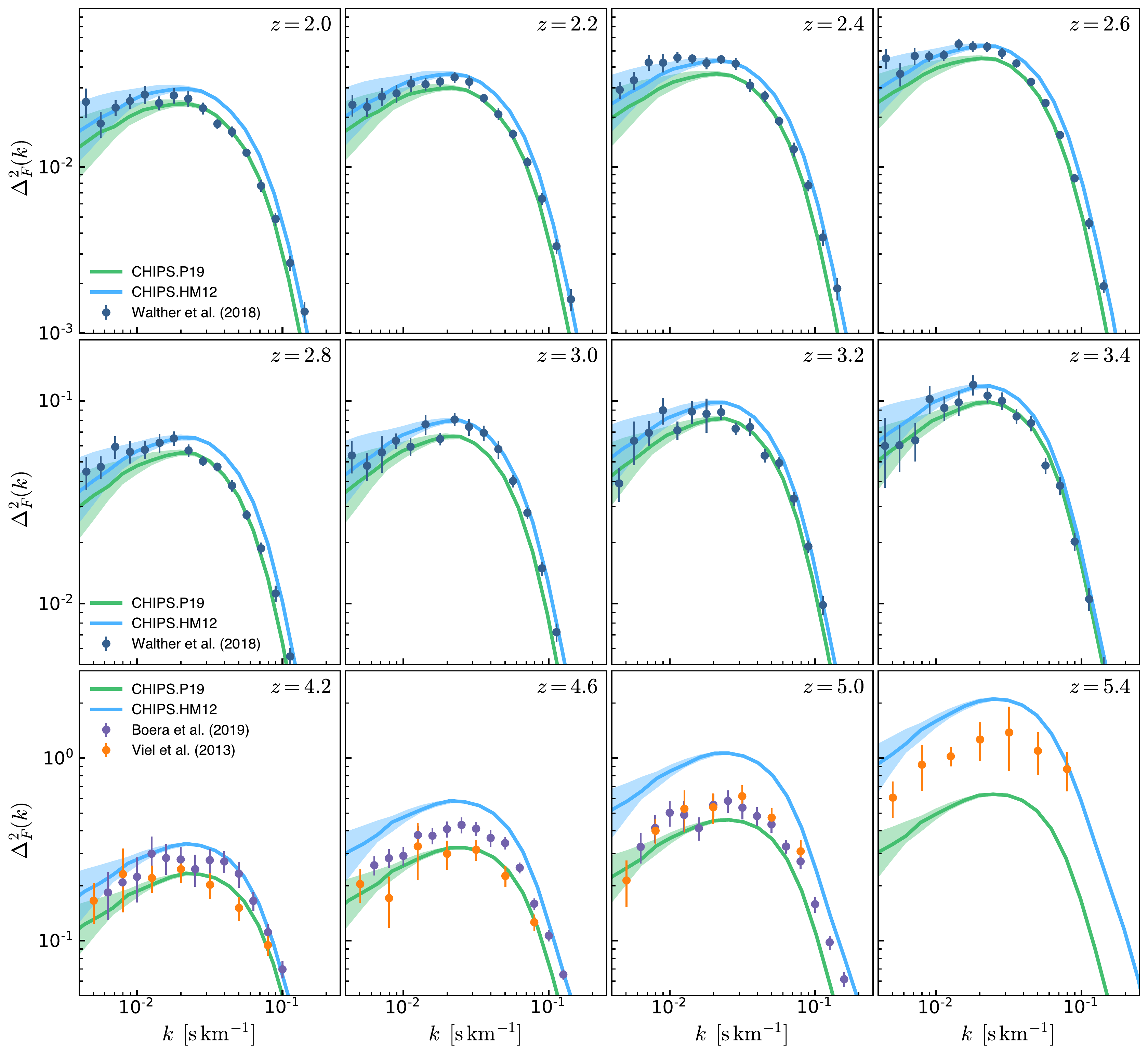}
\caption{One dimensional power spectra of the Lyman-$\alpha$ transmitted flux fluctuations (FPS) from our two simulations \Simn.HM12 (blue) and \Simn.P19 (green), compared with the observational measurements from \cite{walther2018a}, \cite{boera2019a} and \cite{viel2013a} (data points). 
The colored lines show the FPS averaged over all the skewers, and the shaded regions show the $\sigma(k) / \sqrt{N_{\mathrm{ind}}}$ region where $\sigma(k)$ is the standard deviation of the distribution $P(\Delta_F^2)$ obtained from the FPS of the all the individual skewers and $N_{\mathrm{ind}}(k)$ is the number of independent skewers that can be drawn from the simulation grid for each axis. For $k \gtrsim 0.01 \,\,\skm$ the agreement between \Simn.P19 and the observational measurement of $P(k)$ for $ 2 \lesssim  z \lesssim 4.5 $ is relatively good (time-averaged $\langle \chi_\nu^2 \rangle \sim 2$), as compared with \Simn.HM12 ($\langle \chi_\nu^2 \rangle \sim 8$).
For $z \gtrsim 5$ the amplitude of $P(k)$ for the P19 UVB model is lower than the data, reflecting the lower estimate of $\tau_{eff}$ from the P19 model relative to the observations. For $ k \lesssim 0.01 \,\, \skm$ the resulting FPS from the HM12 model is in better agreement with the data from \cite{walther2018a} at $z \lesssim 3$ ($ \langle \chi_\nu^2 \rangle \sim 1$) in contrast to the P19 model ($ \langle \chi_\nu^2 \rangle \sim 3$).}
\label{fig:flux_power_spectrum_0}
\end{figure*}

\begin{figure*}
\includegraphics[width=\textwidth]{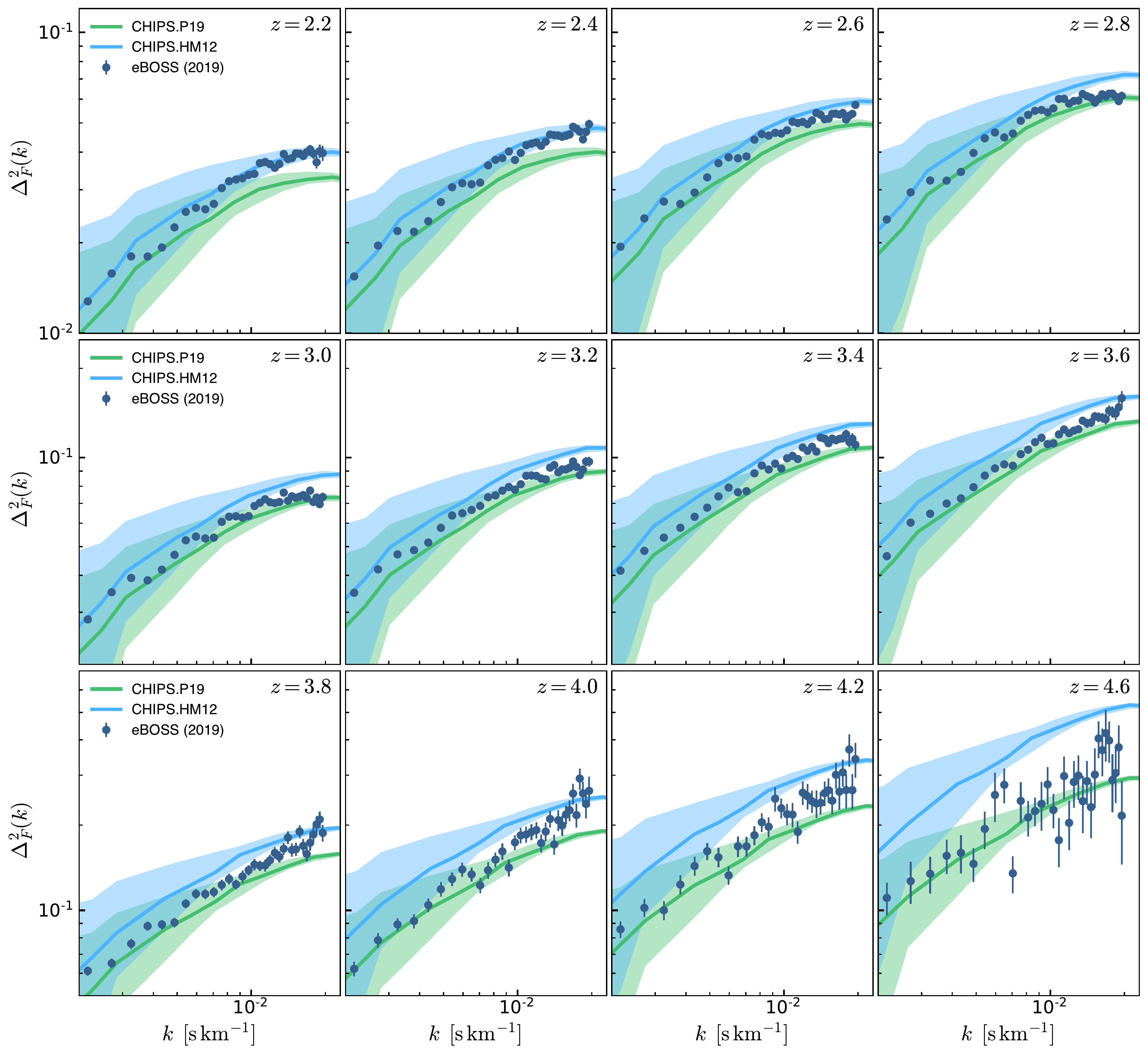}
\caption{One dimensional power spectra of the \Lya transmitted flux fluctuations (FPS) from our two simulations \Simn.HM12 (blue) and \Simn.P19 (green),  compared with the large scale power spectra data from the eBOSS experiment \citep{BOSS_2019}. The colored lines show the FPS averaged over all the skewers, and the shaded regions show the $\sigma(k) / \sqrt{N_{\mathrm{ind}}}$ region where $\sigma(k)$ is the standard deviation of the distribution $P(\Delta_F^2)$ obtained from the FPS of the all the individual skewers and $N_{\mathrm{ind}}(k)$ is the number of independent skewers that can be drawn from the simulation grid for each axis. 
From $z=2.2$ to $z=4.6$ the amplitude of $P(k)$ for the models increases faster than the data, and for the range $ 2.6 \lesssim z \lesssim 4.2 $ the data lie in between the models.
The P19 and HM12 models result in $P(k)$ that match the data at $z = 4.6$ and $z = 2.2$, respectively.}
\label{fig:flux_power_spectrum_1}
\end{figure*}

On scales of a few Mpc, the \Lya flux power spectrum (FPS) is an excellent probe of the thermal properties of the photoionized IGM and can constrain the IGM temperature at various epochs.
On scales below $\sim100$kpc, 
the FPS exhibits a cutoff beyond which the \Lya forest has suppressed
structure owing to both the pressure smoothing of the gas distribution (which sets the pressure smoothing scale $\lambda_P$) as well as the thermal Doppler broadening of the absorption lines.
To compare the FPS produced by the UVB models in the simulations with observation, we measured the \Lya transmitted flux power spectra from the large sample of skewers computed as described in \S \ref{sec:synthetic_spectra_skewer}.
The FPS as a function of velocity $u$ is obtained from  the flux fluctuations as
\begin{equation}
\delta_{F}(u) \equiv \frac{F(u)-\bar{F}}{\bar{F}},
\end{equation}
\noindent
where $F(u)$ if the transmitted flux along the skewer over the velocity interval [0, $u_{max}$]  and $\bar{F} = \exp(-\tau_{eff})$ is the transmitted flux averaged over all the skewers. The FPS is commonly expressed in terms of the dimensionless quantity $\Delta^2_{F}$, defined as
\begin{equation}
\Delta_{F}^{2}(k)=\frac{1}{\pi} k P(k).
\end{equation}
\noindent
The transmitted flux power spectrum $P(k)$ is computed as
\begin{align}
P(k)&=u_{\mathrm{max}}\left\langle\left|\tilde{\delta}_{F}(k)\right|^{2}\right\rangle \\
\tilde{\delta}_{F}(k)&=\frac{1}{u_{\mathrm{max}}} \int_{0}^{u_{max}}  e^{-i k u} \delta_{F}(u) d u,
\end{align}
\noindent
where $k=2\pi/u$ corresponds to the wavenumber associated to the velocity $u$ and has units of $\skm$ and $u_{\mathrm{max}}$ is the Hubble flow velocity for the box length $u_{\mathrm{max}} = H L / ( 1 + z) $. From our simulations,
we measured the FPS resulting from both UVB models and compared with the analogous observational measurements.
Our results are shown in Figures \ref{fig:flux_power_spectrum_0} and \ref{fig:flux_power_spectrum_1},
where the lines show the FPS averaged over all the skewers and the shaded bars show the $\sigma(k) / \sqrt{N_{\mathrm{ind}}(k)}$ region. Here $\sigma(k)$ corresponds to the standard deviation from the distribution of  $P(k)$ measured over all individual skewers and  $N_{\mathrm{ind}}(k) = (2 \pi)^{-2} ( k  u_{max} )^2 $ is the number of independent skewers that can be drawn from the simulation grid for each axis. Applying the Central Limit Theorem, $\sigma(k) / \sqrt{N_{\mathrm{ind}}(k)}$ is analogous to the standard deviation $\sigma_s(k)$ of the distribution that results from sampling the mean FPS ($\overline{P(k)}$) over all the possible groups of independent skewers, therefore  $\sigma_s(k) \simeq \sigma(k) / \sqrt{N_{\mathrm{ind}}(k)}$ provides the uncertainty in the mean FPS measured from the simulation grid.

The flux power spectrum  on scales of $0.004 <  k < 0.2  \,\,  \skm$ is presented in Figure \ref{fig:flux_power_spectrum_0}, as shown, the uncertainty bars in the in the power spectrum from the simulations are larger for smaller $k$ values as the number of independent skewers that can be drawn from the grid decreases as the size of the probed fluctuation increases. We compare the simulation results against observational measurements from \cite{walther2018a} at $2<z<3.4$,
and higher redshift measurements from \cite{boera2019a} and \cite{hiss2018a} for $4\lesssim z \leq 5.4$. To assess the performance of both photoionization and photoheating models to reproduce the observed FPS, we quantify the differences in $P(k)$ 
with the statistic $\chi^2_\nu = \chi^2 / N$, where
\begin{equation}
\label{eqn:chi2}
\chi^{2} = \sum_{i}^N \left[ \frac{ P(k_i)^{\mathrm{obs}} - P(k_i)^{\mathrm{model}} } { \sigma_i^{\mathrm{obs}} }  \right]^{2}, 
\end{equation}   
\noindent
and $N$ is the number of observed data points measured at the wavenumbers $k_i$ and having uncertainties $\sigma_i$.
We will use $\chi^2$ to denote the statistic computed by Equation \ref{eqn:chi2} for multiple scales at the same
redshift, and $\langle \chi^2 \rangle$ to denote a ``time-averaged'' statistic when using multiple scales over multiple redshifts.

For $k \gtrsim 0.01 \,\, \skm$, the agreement between the \Simn.P19  simulation (green)
and the observational measurement of $P(k)$ for $ 2 \lesssim  z \lesssim 4.5 $ is relatively good as the time-averaged differences are $\langle \chi_\nu^2 \rangle \sim 2$,  compared to $\langle \chi_\nu^2 \rangle \sim 8$ for \Simn.HM12 (blue). At high redshifts ($z \gtrsim 5 $),
the observational data lie between the predictions from the two models. This result is
consistent with the behavior of $\tau_{eff}$ in Figure \ref{fig:optical_depth_redshift},
since the normalization of the transmitted flux fluctuations $\delta F$ is determined by $\bar{F}$
and
an underestimate of $\tau_{eff}$ will result in a lower normalization in $P(k)$.
For $ k \lesssim 0.01 \,\,  \skm$, the HM12 UVB model produces FPS that agree better
with
the data from \cite{walther2018a} at $z \lesssim 3$ ($ \langle \chi_\nu^2 \rangle \sim 1$) than the P19 model ($ \langle \chi_\nu^2 \rangle \sim 3$).
At high redshift ($z \gtrsim 4.6$),
the HM12 model results in a FPS higher than the observations, and
this discrepancy is again consistent
with the higher values of $\tau_{eff}$ produced by the model compared with the observations, 
as shown in Figure \ref{fig:optical_depth_redshift}.

For larger scales $0.002 < k < 0.02 \,\,  \skm$, the flux power spectrum shown in Figure \ref{fig:flux_power_spectrum_1} 
is compared with the observational measurements from the eBOSS experiment presented in \cite{BOSS_2019}.
The evolution of $P(k)$
in \Simn.HM12 (blue) and \Simn.P19 (green) differs from the observed data, with
the amplitude of $P(k)$ in the models increasing faster than the data at $z=2.2-4.6$.
Figure \ref{fig:flux_power_spectrum_1} shows that at higher redshift ($z \sim 4.6$)
the P19 model $P(k)$ matches the observations ($\chi_\nu^2 = 1.4$),
while for the redshift range $ 2.6 \lesssim z \lesssim 4.2 $ 
the data lies in between the models.
For $z \lesssim 2.4$, the HM12 model agrees better with the data measured by the eBOSS experiment but the small uncertainties in the observational measurements result in large values of $\chi_\nu^2$ regardless.
Since the temperature of the IGM at $  z \lesssim 4$  is primarily set by \HeII reionization,
we argue that the discrepancies between the P19 results and the observed data could be
alleviated by changing the \HeII photoheating rate associated with active galactic nuclei
to reduce the IGM temperature at $z\sim3$.

Comparing \Lya flux power spectrum between simulations and observation offers a
direct way to assess the performance of the chosen photoionization and photoheating rates 
in reconstructing the thermal history of the IGM.
Figures \ref{fig:flux_power_spectrum_0} and \ref{fig:flux_power_spectrum_1} show that
both photoionization and photoheating models used for this work, \cite{haardt2012a} and \cite{puchwein2019a},
fail to recover the observed $P(k)$ on scales of $ 0.002 < k < 0.2 \,\,  \skm$ over the redshift range $ 2 \lesssim z \lesssim 5$.
The observed \Lya forest statistics $\tau_{eff}$ and $P(k)$ lie in between the results produced by the two models.
This tension motivates further studies using cosmological simulations 
with modified photoionization and photoheating rates that
result in lower IGM temperatures at $ 2 < z \lesssim 3$ and
increase the amplitude of the FPS on large scales (e.g., $0.002 \lesssim k \lesssim 0.02 \,\,  \skm$)
to match better the observations.

\section{Discussion}
\label{sec:discussion}

By comparing the results of our simulations with observations, we have demonstrated
the broad properties of the forest are reproduced by models using either
the \citet{haardt2012a} or \citet{puchwein2019a} photoheating and photoionization rates.
The \citet{puchwein2019a} rates in particular lead to realistic small-scale structure in the
forest over a range in redshift.
However, both models fail to recover the detailed shape of the transmitted
flux $P(k)$ or the magnitude of the optical depth 
at all redshifts or spatial scales. The physical reasons for these inadequacies
likely also depend on redshift and spatial scale, and are discussed below.

\subsection{What is the IGM Photoheating History?}
\label{sec:model_discussion}

As Figures \ref{fig:thermal_history} and \ref{fig:thermal_history_data}
illustrate, the observational inferences on the evolution of the
IGM  mean temperature and density-temperature relation are currently
widely-varying
\citep{hiss2018a,bolton2014a,walther2018a,boera2019a,gaikwad2020a,gaikwad2020b}.
Deriving these observed properties requires assistance from simulations,
for instance by generating model spectra or aiding
the interpretation of transmission spikes. As a result,
discriminating between early-reionization \citep{haardt2012a} and  late-reionization \citep{puchwein2019a}
based on $T_0(z)$ and $\gamma(z)$ remains hazardous. The
\Lya optical depth and $P(k)$ evolution show that the observations
mostly reside in between the model predictions. At redshifts 
$z\sim2-3$ the \citet{puchwein2019a} rates produce structure in the
forest that agrees better with the data on small scales, but on
larger scales \citet{haardt2012a} performs better. The IGM
structure at these redshifts is heavily influenced by \ion{He}{2}
photoheating powered by active galactic nuclei, and both 
the relative
spectral slopes of the AGN spectral energy distribution and
the difference in emissivity with redshift in these
models could affect their scale-dependent relative agreement.
Finding the \ion{He}{2} photoheating and photoionization rates that result
in agreement across all scales at these redshifts will require
future work and more simulations.

Close to the hydrogen reionization era,
in simulations using the \citet{haardt2012a} rates the
\Lya $P(k)$ amplitude increases more rapidly with redshift than in the
\citet{puchwein2019a} model. Reionization occurs very early in the
\citet{haardt2012a} UVB, and the $P(k)$ amplitude evolution reflects
the progressively colder IGM in this scenario. The
hydrogen
ionization and temperature states of the IGM in the \citet{puchwein2019a}
model are apparently too high, and produce a lower $P(k)$ than seen
in the observation. Balancing the high temperature resulting from the
late reionization in this model and the larger $P(k)$ may require 
a patchy reionization process, as noted previously \citep{keating2020a}.

\subsection{IGM Thermal History vs. Instantaneous Properties}
\label{sec:history}

The importance of developing self-consistent histories for
IGM properties, including the phase structure, flux power spectra,
and \Lya optical depth, cannot be overstated for interpreting
observations. The power of the \Lya forest for constraining
the small-scale physics of structure formation, including the
possible presence of warm dark matter \citep[e.g.,][]{viel2013a,irsic2017a}
and the importance of
neutrinos \citep[e.g.,][]{BOSS_2019}, is limited by the imprecisely known
thermal properties that can impact such scales.
These uncertainties on the thermal properties, often characterized
by the temperature $T_0$ at the mean density and the slope $\gamma$
of the IGM temperature-density relation, have frequently
been treated as nuisance parameters when developing cosmological 
parameter constraints from the \Lya forest. Analyses typically marginalize
over the uncertainties in the thermal structure of the forest to
arrive at, e.g., the possible contribution of warm dark matter
to the suppression of small-scale power.

One complication with these analyses that
our simulations highlight is how the history and evolution
of the thermal properties influence the structure of the forest.
The characteristics of the forest at one redshift cannot
be disentangled entirely from its properties at similar redshifts.
The response of the forest, in terms of both its thermal and ionization
structure, depends on the evolving photoheating and photoionization
rates, and the values of both $T_0$ and $\gamma$ change along
tracks with redshift. These properties are not independent, and
have redshift correlations that cannot be ignored by separately
marginalizing over their properties independently at an 
array of redshifts. Instead, when marginalizing over the
thermal structure of the forest to infer constraints from 
$P(k)$, full simulated histories of the forest properties
are required with marginalization occurring simultaneously over
the forest structure at all redshifts where observations are
available. For instance, synthesizing the results shown in 
Figures \ref{fig:thermal_history}, \ref{fig:thermal_history_data},
\ref{fig:flux_power_spectrum_0}, we find that the consistently
lower $P(k)$ amplitude of the forest at $z=[4.2,4.6,5.0,5.4]$
when using the \citet{puchwein2019a} rates results from the
hotter IGM induced by the late global reionization in this model. The
temperature at these times is not independent across redshift,
and varying models over a range of IGM temperature and \Lya $P(k)$
at a given
redshift amounts to direct assumptions on those properties at
adjacent redshifts where the dominant photoheating mechanisms
are the same.

Addressing this issue requires a potentially large number of
hydrodynamical simulations of cosmological structure formation
that capture various photoheating histories and manage to resolve
the \Lya forest structure robustly on small scales. Our 
\Sim simulations represent a first step in this direction, and
the computational efficiency of the \Cholla code will enable
us to realize the required number of simulations with moderate
additional effort.

\section{Summary}
\label{sec:summary}

Motivated by new observational efforts that will provide unprecedented
detail on the properties of the \Lya forest \citep[e.g.,][]{desi2016a},
we have initiated the \Sim series of hydrodynamical simulations of
cosmological structure formation. Our simulations use the GPU-native
\Cholla code to maintain exquisite spatial resolution in the low-density
intergalactic medium throughout cosmological volumes. In this first
paper, we conduct $N=2048^3$ resolution simulations to compare
the thermal history and physical properties of 
the \Lya forest using two models for the photoheating and
photoionization rates induced by an evolving ultraviolet background
\citep{haardt2012a,puchwein2019a}. A summary of our efforts and
conclusions follows.

\begin{itemize}
	\item We extended the \Cholla code to perform cosmological simulations
	by engineering gas self-gravity, a Fourier-space Poisson solver, a
	particle integrator, a comoving coordinate scheme, and a coupling
	to the GRACKLE heating and cooling library \citep{smith2017a}.
	
	\item We provided extensive tests of our cosmological simulations,
	including the \citet{zeldovich1970a} pancake test and comparisons
	with the results of other cosmological simulation codes. \Cholla agrees
	with \Nyx \citep{almgren2013a}, \Ramses \citep{teyssier2002a},
	and \Enzo \citep{bryan2014a} to sub-percent accuracy on all spatial
	scales for N-body cosmological simulations, and matches \Ramses
	and \Enzo results to within a few percent for adiabatic hydrodynamical
	simulations.
	
	\item We provide a new method for testing the dual energy formalism
	\citep{bryan1995a} of Eulerian hydrodynamical codes for cosmological 
	simulations by matching the mean cosmic gas temperature in adiabatic 
	simulations, and show that \Cholla recovers the expected results.
	
	\item In accordance with prior results \citep[e.g.,][]{puchwein2019a},
	we find that after hydrogen reionization, simulations using the 
	\citet{haardt2012a} photoheating rates predict a cooler IGM temperature
	than the \citet{puchwein2019a} UVB model. At redshfits $z\sim4-6$,
	the \citet{puchwein2019a} model is hotter owing to hydrogen reionization
	completing later than in the \citet{haardt2012a} scenario. At redshifts
	$z\lesssim4$, the IGM is hotter in the \citet{puchwein2019a} model owing
	to the \ion{He}{2} photoheating rates powered by active galactic nuclei.
	
	\item We compare the \Lya transmitted flux power spectra $P(k)$ computed
	for these simulations with observations.
	We find that at redshifts $2\lesssim z \lesssim 5.5$ the performance of the
	models varies with scale. Using the \citet{puchwein2019a} photoheating rates
	results in good agreement with the observed $P(k)$ on $k\gtrsim0.01~\skm$
	at $2\lesssim z \lesssim 4.5$.
	
	\item On larger scales, the amplitude of the observed $P(k)$
	increases faster from  $z\sim2.2$ to $z\sim4.6$ than the structure in the
	simulated forest. The observations appear intermediate between the 
	simulation results using the \citet{haardt2012a} and \citet{puchwein2019a} photoheating
	rates at $2.2\lesssim z \lesssim 4.6$ for $k\approx 0.002-0.01~\skm$.
	
	\item At higher redshifts $z\gtrsim4.5$, as the epoch of hydrogen
	reionization is approached, the $P(k)$ amplitude in the simulations increase at
	rates that bracket the observed flux $P(k)$. The observed \Lya optical depth
	also lies in between the model predictions at these redshifts.
	
	\item We show that our results
	are insensitive to small changes in the cosmological parameters comparable
	to the \citet{planck2018a} uncertainties, and demonstrate our results for
	the flux power spectra have converged with resolution studies.
\end{itemize}

These initial \Sim simulations demonstrate that commonly used models for the
photoheating and photionization rates \citep{haardt2012a, puchwein2019a} broadly
reproduce the observed thermal history and transmitted flux power spectra of
the \Lya forest. However, in detail the agreement with the observations can
be improved, including better recovering the redshift- and scale-dependence of the
the flux $P(k)$ and the evolution in the \Lya optical depth. Matching these
observations more completely will require changing the photoionization and
photoheating rates for both hydrogen and helium. We will explore these
improvements using additional large-scale cosmological simulations in future work.

\acknowledgments

We wish to thank Francesco Haardt and Ewald Puchwein for their discussions and guidance.
The simulations used for this paper were produced in the Summit system (ORNL) project AST149. 
This research used resources of the Oak Ridge Leadership Computing Facility at the Oak Ridge National Laboratory, which is supported by the Office of Science of the U.S. Department of Energy under Contract No. DE-AC05-00OR22725.
We acknowledge use of the \emph{lux} supercomputer at UC Santa Cruz, funded by NSF MRI grant AST 1828315. 
Bruno Villasenor is supported in
part by the UC MEXUS-CONACyT doctoral fellowship. BER acknowledges support from NASA contract NNG16PJ25C
and grant 80NSSC18K0563. We acknowledge the comments and suggestions received from the paper referee which helped improve the content and clarity of this work. 

\smallskip
\smallskip

\textit{Software:} \Cholla \citep[\url{https://github.com/cholla-hydro/cholla}]{schneider2015a}, Python \citep{Python}, Numpy \citep{numpy}, 
Matplotlib \citep{matplotlib}, MUSIC \citep{MUSIC}, PFFT \citep{PFFT}, GRACKLE \citep{smith2017a}, CLOUDY \citep{Cloudy2017}, ROCKSTAR \citep{Behroozi+2013}.
 
\appendix

\section{Resolution Convergence Analysis}
\label{sec:resolution}

To assess the numerical convergence of our results,
we performed runs with different resolutions.
Each simulation was run with the same box size ($L=50 h^{-1}$Mpc) and identical
cosmological parameters \citep{planck2018a}, but with differing resolutions of
$N=512^3$, $N=1024^3$, and $N=2048^3$ cells and dark matter particles.
These simulations have comoving spatial resolutions of $\Delta x = 98$, 49, and 24 $h^{-1}~\mathrm{kpc}$,
respectively. The initial conditions were generated to preserve the
large-scale modes in common to each simulation, such that the properties of the simulations could be compared
directly on shared spatial scales.
We measured the \Lya effective optical depth and transmitted flux power spectrum at each resolution.

\begin{figure*}
\includegraphics[width=\textwidth]{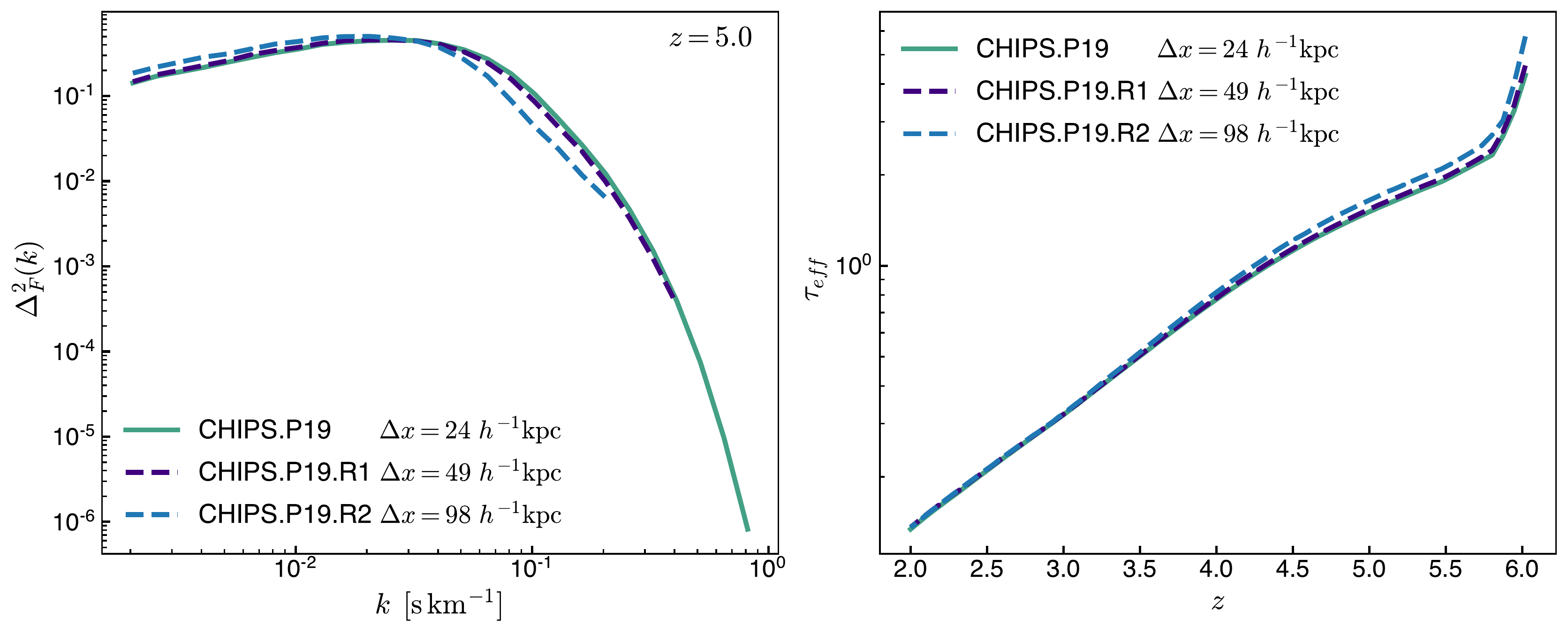}
\caption{\Lya forest statistics measured for similar simulations with different spatial resolutions $\Delta x = 98$, 49 and 24 $h^{-1}\mathrm{kpc}$ (comoving). All simulations have $L= 50 h^{-1}\mathrm{Mpc}$ and use the \cite{puchwein2019a} UVB model and \cite{planck2018a} cosmological parameters. The transmitted flux power spectrum $\Delta_F^2$ at $z=5$ is shown in the left panel and the redshift evolution of the effective optical depth $\tau_{eff}$ is shown on the right. Both measurements demonstrate that the relevant statistics of the \Lya forest have converged for the high resolution simulations used in this work.}
\label{fig:resolution}
\end{figure*}

The left panel of Figure \ref{fig:resolution} shows the convergence of the
\Lya transmitted flux power spectra with resolution.
As the resolution increases, the large-scale $P(k)$ 
decreases while the small-scale power increases. This progression reflects the structure of the forest
becoming better resolved as the number of cells increases. The $N=1024^3$ and $N=2048^3$ simulations agree well over most spatial scales, with little evidence that the $N=2048^3$ simulation requires further refinement on scales $k\gtrsim0.007~\skm$.

The right panel of 
Figure \ref{fig:resolution} details the redshift evolution of the \Lya optical depth measured in the
convergence study. As the structure of the forest becomes better resolved, the optical depth
lowers. This decline reflects the decrease in large-scale power as the resolution improves, as illustrated in the left panel of 
Figure \ref{fig:power_spectrum_comparison}. As with the $P(k)$ analysis, the $N=1024^3$ and $N=2048^3$ simulations agree well
in their \Lya optical depth evolution. The large-scale structure of the forest also agrees well between
these simulations, demonstrating that the $N=2048^3$ simulations have converged on scales that most
contribute to the optical depth.

\section{Cosmological Parameter Study}
\label{sec:parameter_study}

\begin{deluxetable}{lcc}
\tablenum{2}
\caption{Mean Gas Density Comparison for Alternative Cosmologies\label{tab:gas_dens}}
\tablehead{
	\colhead{Simulation} & \colhead{Mean baryon density } & \colhead{$\Delta \bar{\rho_b}/\bar{\rho_b}$ }\\[-6pt]
	\colhead{} & \colhead{$\bar{\rho_b}$  $[\,\mathrm{M_{\odot}}\mathrm{kpc}^{-3}\,]$}  & \colhead{$[\times 10^{-2}]$}
}
\startdata
\Simn.P19 &   6.315 &  \\
\hline
\Simn.P19.A1 & 6.275 & -0.62 \\
\Simn.P19.A2 & 6.334 & 0.31  \\
\Simn.P19.A3 & 6.394 & 1.25  \\
\Simn.P19.A4 & 6.449 & 2.12  \\
\enddata
\end{deluxetable}

\begin{figure}
\centering
\includegraphics[width=0.5\textwidth]{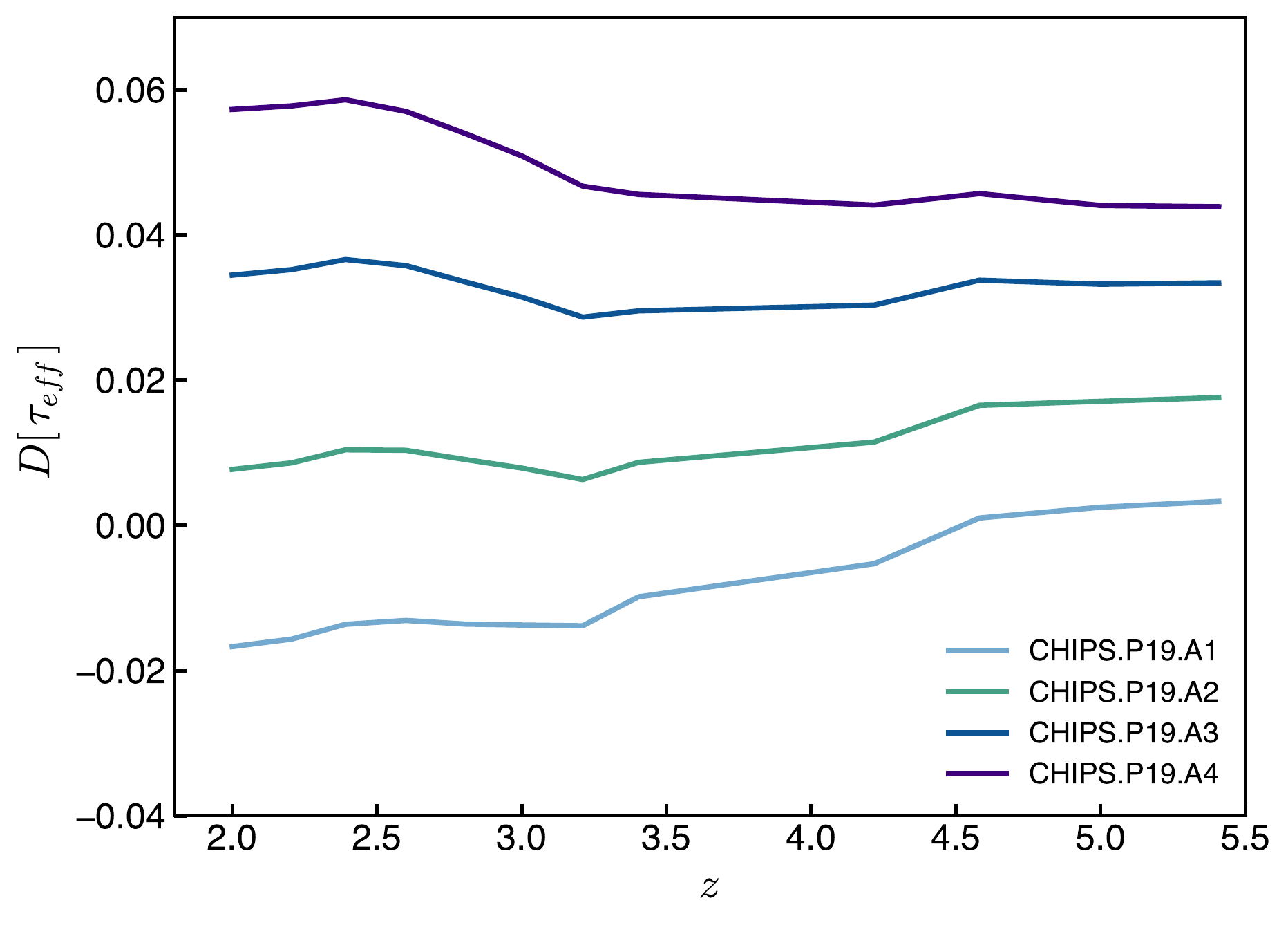}
\caption{Redshift-dependent fractional difference of the effective optical depth $\tau_{eff}$ in the alternative cosmology simulations, measured with respect to the fiducial \Simn.P19 simulation that evolves a \cite{planck2018a} cosmology. The variations in $\tau_{eff}$ reflect differences in the mean baryonic density $\bar{\rho_b}$ for the alternative cosmologies shown in Table \ref{tab:gas_dens}.} 
\label{fig:tau_comparison}
\end{figure}

\begin{figure*}
\includegraphics[width=\textwidth]{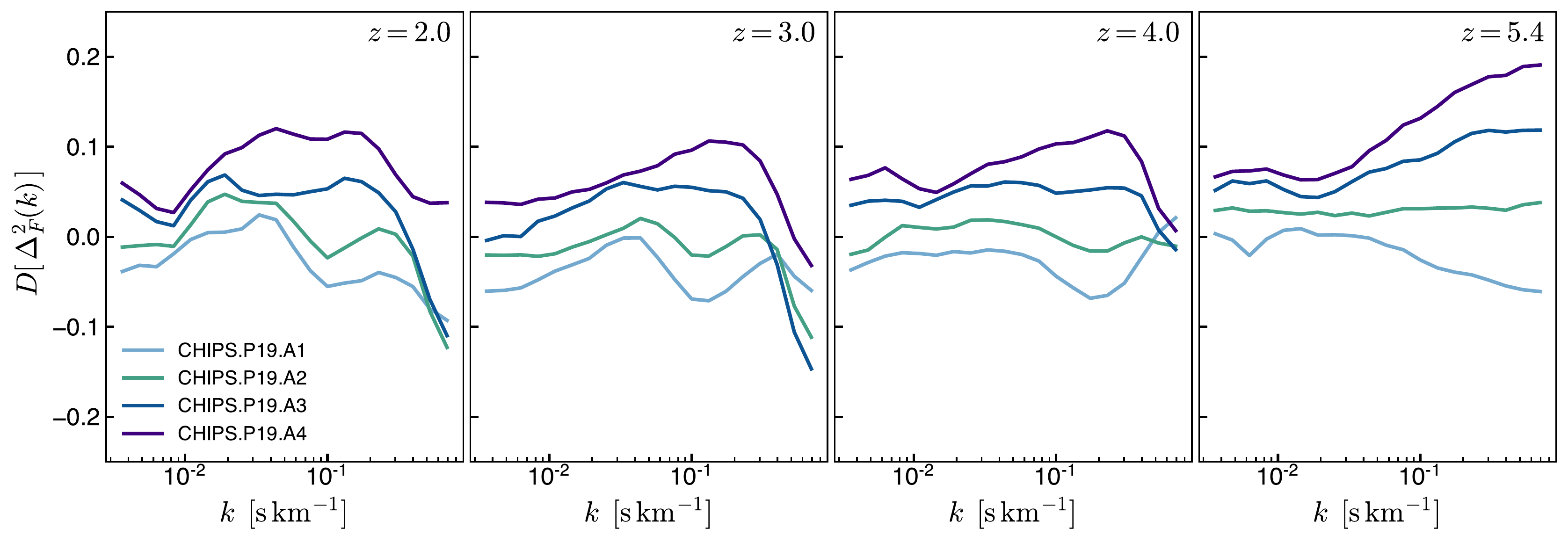}
\caption{Fractional differences in the flux power spectra $\Delta_F^2$ of the alternative cosmology simulations. The
differences are measured with respect to the fiducial \Simn.P19 simulation that evolves a \cite{planck2018a} cosmology (shown for redshifts $z=[2, 3, 4, 5.4]$). The differences are scale dependent, and the
overall variations in normalization reflect the differences in $\tau_{eff}$ shown in Figure \ref{fig:tau_comparison}. }
\label{fig:power_spectrum_comparison}
\end{figure*}

The relative difference in the \Lya forest properties between simulations using the
 \citet{haardt2012a} or \citet{puchwein2019a} rates is substantial, but understanding
 whether these differences are large compared differences in the forest resulting from
 cosmological parameter variations requires further simulation. To answer this question
 we repeated our highest resolution simulation using the \citet{puchwein2019a} rates
 with a range of cosmological parameters, varying over the reported uncertainty in
 the \citet{planck2018a} analysis. We ran four simulations
 where we varied the Hubble parameter $H_0$, 
 matter density $\Omega_m$, baryon density $\Omega_b$, RMS fluctuations on $8~h^{-1}\mathrm{Mpc}$ scales
 $\sigma_8$, and the spectral slope of initial perturbations $n_s$. The chosen numerical
 values of the cosmological parameters are reported in Table \ref{tab:sims}. 
 
Table \ref{tab:gas_dens} shows the mean comoving baryonic density $\bar{\rho_b}$ for the reference simulation \Simn.P19 and the simulations with alternative cosmologies.
Additionally, 
for the alternative cosmologies the fractional difference of the average density $\Delta \bar{\rho_b}/\bar{\rho_b}$ relative to the reference \Simn.P19 is shown in the second column. The differences in the mean gas density are 
approximately a few percent and reflect differences in the mean optical depth measured in the alternative cosmology simulations with respect to the \Simn.P19 simulation.
The differences in the effective optical depth $D[\tau_{eff}]$ relative to \Simn.P19 as a function of redshift are shown in Figure \ref{fig:tau_comparison}.
The differences in $\tau_{eff}$ range from -2\% to 6\%, exhibit little evolution with redshift,
and cannot, e.g., account for the deviations of $\tau_{eff}$ resulting from the \cite{puchwein2019a} model and the observational measurements shown in Figure \ref{fig:optical_depth_redshift}.

Figure \ref{fig:power_spectrum_comparison} shows the resulting
variation in the transmitted flux $P(k)$ induced by the small variations introduced in the cosmological parameters. The differences in the power spectrum $D[\Delta_F^2]$ relative to the reference \Simn.P19 are shown for four snapshots at redshifts between $z=2.0$ and $z=5.4$.
The differences in $P(k)$ are scale dependent, but are consistent with the differences shown in Figure \ref{fig:tau_comparison} for $\tau_{eff}$ as the mean transmitted flux $\bar{F}$ sets the normalization of $P(k)$.
To disentangle the effect that varying each cosmological parameter has on the \Lya statistics a more extensive study would have to be performed,  but we conclude the
overall effect of the cosmological variations is small and as Figure \ref{fig:power_spectrum_comparison}
demonstrates, the difference in physical structure between simulations conducted
with the \citet{haardt2012a} or \citet{puchwein2019a} rates cannot be easily mimicked
with cosmological parameter variations allowed by experimental constraints. This result also
emphasizes the need to account for the IGM thermal history when inferring
cosmological properties from the forest.

We also compared the evolution of the thermal parameters $T_0$ and $\gamma$ from the alternative cosmology simulations to the reference \Simn.P19  simulation, and found that the differences in $T_0$ and $\gamma$ relative to the \cite{planck2018a} cosmology were $\lesssim 3 \%$.
Since these differences are of the order of the uncertainties resulting from modeling the density-temperature distribution as a power-law relation (Equation \ref{eq:rho_T_relation}), we do not
report any significant variation in the thermal history of the IGM owing
to small variations of the cosmological parameters.

\bibliography{references}

\end{document}